\documentclass[12pt]{article}
\usepackage{a4wide,epsfig,scalefnt}
\newcommand{\lmc}{l_{m_c}}

\newcommand{\lnsmu}{L_{s\mu}}
\newcommand{\lnsMc}{L_{sM_c}}

\newcommand{\grtsim}{\mbox{\raisebox{-3pt}{$\stackrel{>}{\sim}$}}}
\newcommand{\lessim}{\mbox{\raisebox{-3pt}{$\stackrel{<}{\sim}$}}}

\begin{document}    


\title{\vskip-3cm{\baselineskip14pt
\centerline{\normalsize\hfill SFB/CPP-07-03}
\centerline{\normalsize\hfill TTP07--02}
\centerline{\normalsize\hfill hep-ph/0702103}
\centerline{\normalsize\hfill February 2007}
}
\vskip.7cm
Heavy Quark Masses from Sum Rules\\ in Four-Loop Approximation
}

\author{
{Johann H. K\"uhn}$^{a}$,
{Matthias Steinhauser}$^a$,
{Christian Sturm}$^b$
  \\[3em]
  {\normalsize (a) Institut f\"ur Theoretische Teilchenphysik,}\\
  {\normalsize Universit\"at Karlsruhe, D-76128 Karlsruhe, Germany}
  \\[.5em]
    {\normalsize (b) Dipartimento di Fisica Teorica,}\\ 
    {\normalsize Universit{\`a} di Torino, Italy \& 
      INFN, Sezione di Torino, Italy}
}
\date{}
\maketitle

\begin{abstract}
\noindent
New data for the total cross section
$\sigma(e^+e^-\to\mbox{hadrons})$ in the charm and bottom threshold
region are combined with an improved theoretical analysis, which
includes recent four-loop calculations, to determine the
short distance $\overline{\rm MS}$ charm and bottom quark masses. A detailed
discussion of the theoretical and experimental uncertainties is
presented. The final result for the  $\overline{\rm MS}$-masses,
$m_c(3~\mbox{GeV})=0.986(13)$~GeV 
and
$m_b(10~\mbox{GeV})=3.609(25)$~GeV, 
can be translated into
$m_c(m_c)=1.286(13)$~GeV 
and
$m_b(m_b)=4.164(25)$~GeV.
This analysis is consistent with but significantly more precise than a similar
previous study.

\vspace{.5em}
\noindent
PACS numbers: 12.38.-t 14.65.Dw 14.65.Fy

\end{abstract}

\thispagestyle{empty}
\newpage
\setcounter{page}{1}

\renewcommand{\thefootnote}{\arabic{footnote}}
\setcounter{footnote}{0}


\section{Introduction}

The strong coupling constant and the quark masses are the fundamental input
parameters of the theory of strong interaction.
Quark masses are an essential input for the evaluation of weak decay
rates of heavy mesons and for quarkonium spectroscopy. 
Decay rates and branching ratios of a light Higgs boson, suggested by
electroweak precision measurements, depend critically on the masses of
the charm and bottom quarks, $m_c$ and $m_b$. Last not least, 
confronting the predictions for these masses with experiment is an 
important task for all variants of Grand Unified Theories.
To deduce the values in a consistent way from different experimental
investigations and with utmost precision is thus a must for current 
phenomenology.

Let us recapitulate the main ingredients of the sum rule approach which will
be used in the following analysis.
Originally the idea has been suggested for the analysis of the charm 
quark mass by the ITEP group~\cite{Novikov:1977dq} long ago. 
Subsequently the method has been developed further~\cite{Reinders:1985sr}
and frequently applied to the bottom quark. Most of these later
analyses concentrated on using relatively
high moments which are less sensitive to the continuum contribution
and exhibit a very strong quark mass
dependence~\cite{Penin:1998kx,Melnikov:1998ug,Beneke:1999fe,Hoang:2000fm,Pineda:2006gx}.
However, 
this approach requires the proper treatment of the threshold, in part
the resummation of the higher order terms from the Coulombic binding
and a definition of the quark mass adopted
to this situation like the potential- or
$1S$-mass~\cite{Beneke:1998rk,Hoang:1998nz}, which subsequently has to be
converted to the $\overline{\rm MS}$-mass.
The low moments, in contrast, can be directly expressed by the short-distance
mass at a scale around 3~GeV and 10~GeV for charm and bottom, respectively.
The extrapolation to even higher scales, required for a number of
applications is therefore insensitive to the larger corrections which
would appear in the low energy region.

A detailed analysis of $m_c$ and $m_b$, based on sum rules, has been
performed several years ago~\cite{Kuhn:2001dm} 
(see also Ref.~\cite{Kuhn:2002zr}) and lead to 
$m_c(m_c)=1.304(27)$~GeV and $m_b(m_b)=4.191(51)$~GeV. This is still
one of the most precise results presently available.
During the past years new and more precise data for 
$\sigma (e^+e^-\to\mbox{hadrons})$ 
have become available in the low energy region,
in particular for the parameters of the charmonium and bottomonium
resonances. Furthermore, the error in the strong coupling constant
with its present value\footnote{We adopt the central
  value from Ref.~\cite{Bethke:2006ac} but double the
  error such that the interval
  overlaps with the central value given by the PDG~\cite{Yao:2006px}.}
$\alpha_s(M_Z)=0.1189 \pm 0.0020$~\cite{Bethke:2006ac}
(as compared to $\alpha_s(M_Z)=0.118\pm0.003$ for the last analysis)
has been reduced. Last not least, the 
vacuum polarization induced by massive quarks has recently been
computed in four-loop
approximation~\cite{Chetyrkin:2006xg,Boughezal:2006px}; more 
precisely: its first derivative at $q^2=0$ has been evaluated, which
corresponds to the lowest moment of the familiar $R$-ratio.
A fresh look at the determination of the
quark masses based on these new developments is thus appropriate and will be
presented below.

The outline of the paper is as follows:
The basic assumptions of our approach are presented in
Section~\ref{sec::generalities}. 
In Section~\ref{sec::background} we discuss in detail the evaluation of
the background and recall the theory input required for the quark mass
determination.
The measurements in the charm threshold region are discussed and
applied to the determination of the charm quark mass in
Section~\ref{sec::charm}. 
Similar considerations are used in Section~\ref{sec::bottom} 
to obtain the bottom quark mass. Section~\ref{sec::con}
contains our conclusions.


\section{\label{sec::generalities}Generalities}

The extraction of $m_Q$ from low moments of the
cross section $\sigma(e^+e^-\to Q\bar{Q})$ exploits its sharp rise
close to the threshold for open charm and bottom production and the
important contributions from the narrow quarkonium resonances. 
At best the properties of the lowest bottomonium state, $\Upsilon(1S)$, can
be evaluated in perturbative QCD. In general a differential
description of the cross section close to threshold involves
necessarily low scales, of ${\cal O}(\alpha_s m_Q)$ or
${\cal O}(\alpha_s^2 m_Q)$ and even non-perturbative contributions,
arising, e.g., from condensates $\langle G^2 \rangle$, will enter.
In contrast, by evaluating the moments
\begin{eqnarray}
  {\cal M}_n &\equiv& \int \frac{{\rm d}s}{s^{n+1}} R_Q(s)
  \,,
  \label{eq::Mexp}
\end{eqnarray}
with low values of $n$, the long distance contributions are averaged
out and ${\cal M}_n$ involves short distance physics only, with a
characteristic scale of order $E_{\rm threshold}=2 m_Q$.
Through dispersion relations the moments are directly related to
derivatives of the vacuum polarization function at
$q^2=0$,
\begin{eqnarray}
  {\cal M}_n &=& \frac{12\pi^2}{n!}
  \left(\frac{{\rm d}}{{\rm d}q^2}\right)^n
  \Pi_Q(q^2)\Bigg|_{q^2=0}
  \,,
\end{eqnarray}
which can be evaluated in perturbative QCD. From dimensional
considerations one obtains 
$m_Q\sim \left({\cal M}_n\right)^\frac{1}{2n}$
which implies
\begin{eqnarray}
  \frac{\delta m_Q}{m_Q} \sim \frac{1}{2n} 
  \frac{\delta {\cal M}_n}{{\cal M}_n}
  \,,
\end{eqnarray}
for the relative error induced by the experimental uncertainties.
(The theoretical and parametric errors will be described below.)
Given $\delta R/R \approx 1.5 -3\%$, a precision close to
10~MeV for $m_c$ and 20~MeV for $m_b$ seems within
reach. An analysis at this level obviously requires to control a
variety of corrections, e.g., 
contributions from the non-perturbative condensate
in the case of $m_c$. Furthermore,
a careful definition of the heavy quark production cross section per
se is required. We will discuss this now in detail for charm
production, with the generalization to bottom production being obvious.

The narrow charmonium resonances $J/\Psi$, $\Psi(2S)$ and the
higher excitations will obviously contribute to the moments. Open
charm production exhibits a sharp rise, nearly like a step function.
Beyond the $\Psi(3770)$-resonance a few oscillations are observed which 
quickly level out into a fairly flat energy dependence.
Around and above approximately  5~GeV the cross section is well
approximated by perturbative QCD and mass terms can be considered as
small corrections. 
The sensitivity to $m_Q$ is, therefore, concentrated on the small region
from $J/\Psi$ up to approximately 5~GeV.

\begin{figure}[t]
  \begin{center}
    \begin{tabular}{cc}
      \leavevmode
      \epsfxsize=5cm
      \epsffile[203 302 540 507]{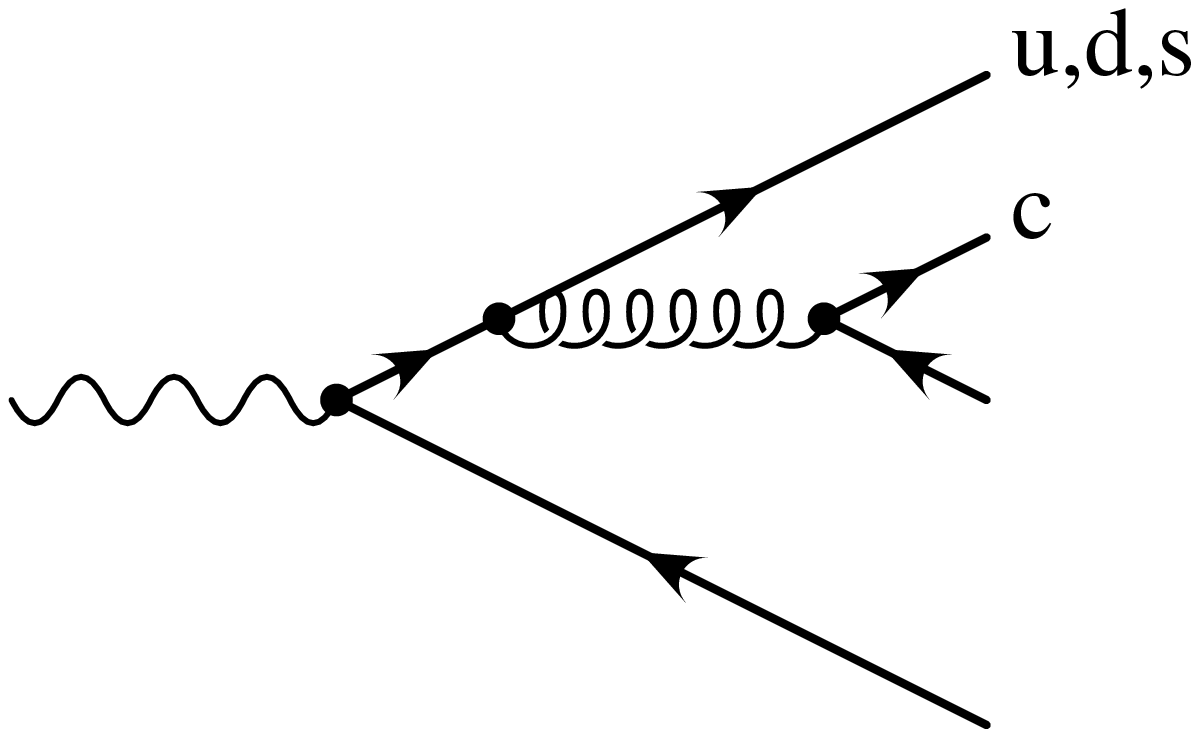}
      &
      \leavevmode
      \epsfxsize=5cm
      \epsffile[203 302 540 507]{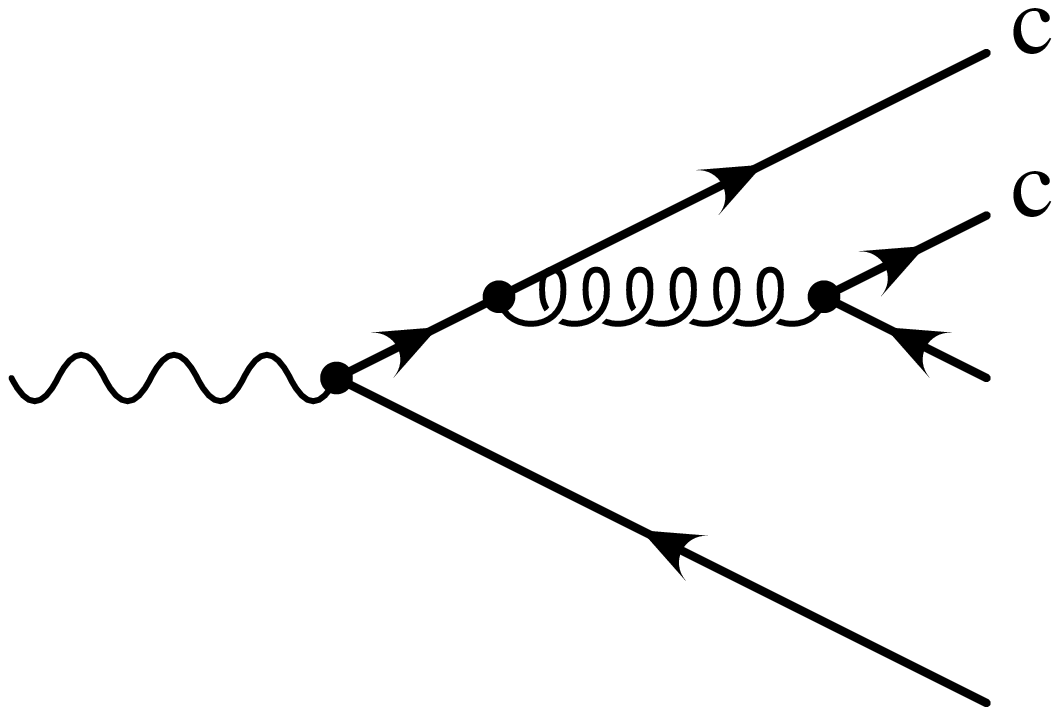}
      \\ (a) & (b)
    \end{tabular}
  \end{center}
  \caption{\label{fig::gacc}Feynman diagrams contributing to $R(s)$ at
    order $\alpha_s^2$. A secondary charm quark pair is 
    produced through gluon splitting.
          }
\end{figure}

\begin{figure}[t]
  \begin{center}
    \begin{tabular}{c}
      \leavevmode
      \epsfxsize=10cm
      \epsffile[60 400 400 600]{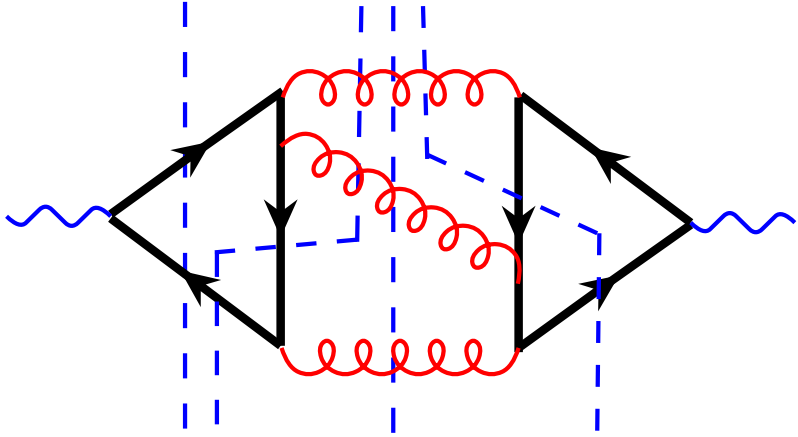}
    \end{tabular}
  \end{center}
  \caption{\label{fig::r_sing}Sample singlet diagram contributing 
    to $\Pi(q^2)$ which 
    arises at order $\alpha_s^3$ for the first time. The cuts
    (indicated by the dashed lines; not all possible cuts are shown) 
    represent contributions to $R(s)$.
          }
\end{figure}

At first glance one might try to extract $\sigma(e^+e^-\to c\bar{c})$ by
measuring the cross section with tagged charmed hadrons, thus eliminating the
contribution from light quarks. However, such
data are presently only available for selected narrow energy regions.
Even more important, 
this approach is also not tenable from the conceptual side.
A sizeable fraction of $\Psi(3770)$ decays 
(not to speak of $J/\psi$ or $\psi'$) proceeds to non-charmed
final states, nevertheless its total contribution should be attributed
to the moments. On the other hand, processes with ``secondary'' charm
production arising from gluon splitting
(cf. Fig.~\ref{fig::gacc}(a)) should rather be assigned to the
light quark cross section. 
Also cuts through singlet diagrams (cf. Fig.~\ref{fig::r_sing}) which
first appear in order $\alpha_s^3$ and where the fermion lines
represent light or heavy quarks,  exhibit thresholds at $q^2=0$,
$4m^2$ and $16m^2$, and these contributions
cannot be attributed to a specific quark flavour. 
For the determination of $m_Q$ we therefore rely on the measurement of
the total cross section, which is calculated through the absorptive
parts of vacuum polarization diagrams. Various contributions to the
cross section, arising from different cuts through the same diagrams,
will always be treated together. For the moments ${\cal M}_n$ 
we only consider those contributions to $\Pi(q^2)$ which are analytic for
$|q^2|\le 4 m_Q^2$ and  exhibit a cut starting at $q^2=4m_Q^2$. 
They result from
diagrams with one massive quark loop connecting both photon vertices
and will be denoted by $\Pi_Q$ and $R_Q$, respectively. The remaining
diagrams with cuts starting at $q^2=0$ will be treated as
``background''. This includes all diagrams with
one light quark loop connecting both photon vertices (including those
with internal massive quark loops)
and all singlet diagrams. It will be shown below that
the background defined in this way exhibits a fairly flat behaviour across the
threshold, despite the fact that individual cuts with branching points
at $q^2=4m_Q^2$ exist.

In the remaining part of this Section we fix the
notation and give precise definitions
of the quantities used in the paper.
It is convenient to normalize the radiatively corrected
hadronic cross section relative to the point cross section 
and to define the ratio
\begin{eqnarray}
  R(s) &=& \frac{\sigma(e^+e^-\to\mbox{hadrons})}{\sigma_{\rm pt}}
  \,,
\end{eqnarray}
where $\sigma_{\rm pt} = 4\pi\alpha^2/(3s)$.
As an inclusive quantity $R(s)$  is conveniently obtained via the optical
theorem from the imaginary part of the 
polarization function of two vector currents via
\begin{eqnarray}
  R(s) &=& 12\pi\,\mbox{Im}\left[ \Pi(q^2=s+i\epsilon) \right]
  \,,
\end{eqnarray}
where $\Pi(q^2)$ is defined through
\begin{eqnarray}
  \left(-q^2g_{\mu\nu}+q_\mu q_\nu\right)\,\Pi(q^2)
  &=&
  i\int {\rm d}x\,e^{iqx}\langle 0|Tj_\mu(x) j^{\dagger}_\nu(0)|0
  \rangle
  \,,
  \label{eq::pivadef}
\end{eqnarray}
with $j_\mu$ being the electromagnetic current.

The perturbative expansion of $R(s)$ can be written as
\begin{eqnarray}
  R(s) &=& \sum_i \sum_n \left(\frac{\alpha_s}{\pi}\right)^n 
  R_i^{(n)}
  + R_{\rm sing}
  \,,
  \label{eq::R}
\end{eqnarray}
where the summation is performed over all active quark flavours $i$.
(The small singlet contribution starts to contribute at order $\alpha_s^3$.)
For a comprehensive compilation of the individual pieces 
we refer to~\cite{Chetyrkin:1996ia,Chetyrkin:1997pn,Harlander:2002ur}
where also explicit results are given. 
The full quark mass dependence is available up to order
$\alpha_s^2$~\cite{Chetyrkin:1995ii,Chetyrkin:1997mb}.
For the non-singlet term $R_i^{(3)}$ the first three terms of the
high-energy expansion are known~\cite{Chetyrkin:2000zk}.

The predictions for $R(s)$ which are used to calculate the
background and the continuum term far above charm or bottom threshold,
respectively, are based on Eq.~(\ref{eq::R}) where the
up, down and strange quark masses are taken as massless.
For the charm and bottom quark the respective pole masses are 
chosen as input.
If not stated otherwise we will use the following parameters
for the evaluation of the background
\begin{eqnarray}
   \alpha_s^{(5)}(M_Z) &=&0.1189\pm0.002\,,
   \nonumber\\
   M_c&=&(1.65\pm0.15)~\mbox{GeV}\,,
   \nonumber\\
   M_b&=&(4.75\pm0.20)~\mbox{GeV}\,,
   \label{eq::input}
\end{eqnarray}
which cover the full range of all currently accepted results.
In Eq.~(\ref{eq::input}) $M_c$ and $M_b$ refer to the pole masses.

At several places of our analysis the renormalization group functions
and the matching conditions for the masses and the strong coupling
are needed in order to get relations between different energy scales.
The corresponding calculations are performed using the package 
{\tt RunDec}~\cite{Chetyrkin:2000yt}.


\section{\label{sec::background}The background}

As stated above, we distinguish three energy regions: 
First, the region of the narrow resonances $J/\psi$ and $\psi'$,
second, the ``charm threshold region''  starting from the $D$-meson threshold
at 3.73~GeV up to approximately 5~GeV, where the cross section exhibits rapid
variations and, third, the continuum region where pQCD and local duality are
expected to give reliable predictions. In this last region the cross section 
is mainly sensitive to $\alpha_s$ and fairly insensitive to $m_c$.

\begin{figure}[t]
  \begin{center}
    \begin{tabular}{c}
      \leavevmode
      \epsfxsize=.8\textwidth
      \epsffile{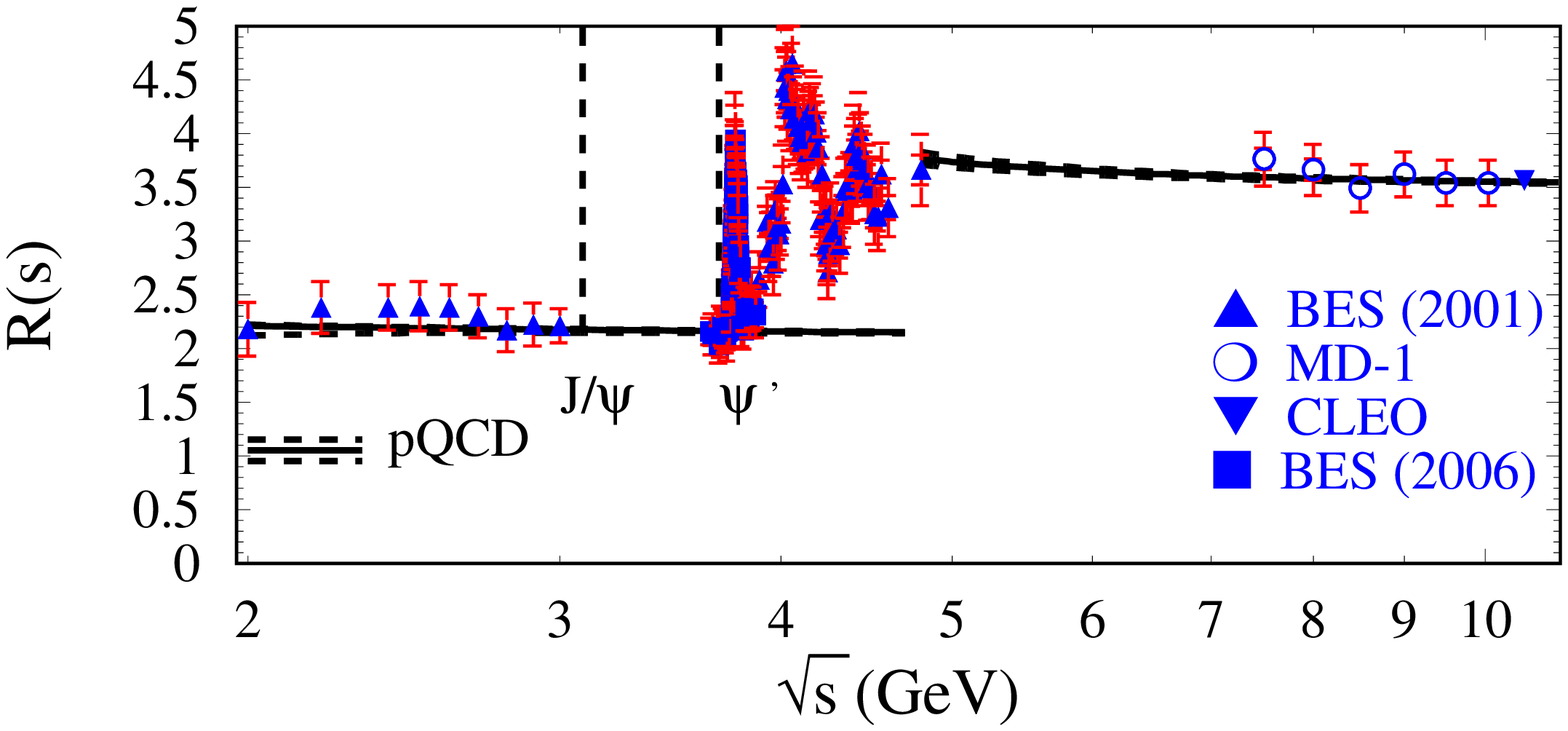}
      \\[-2em]
      \leavevmode
      \epsfxsize=.8\textwidth
      \epsffile{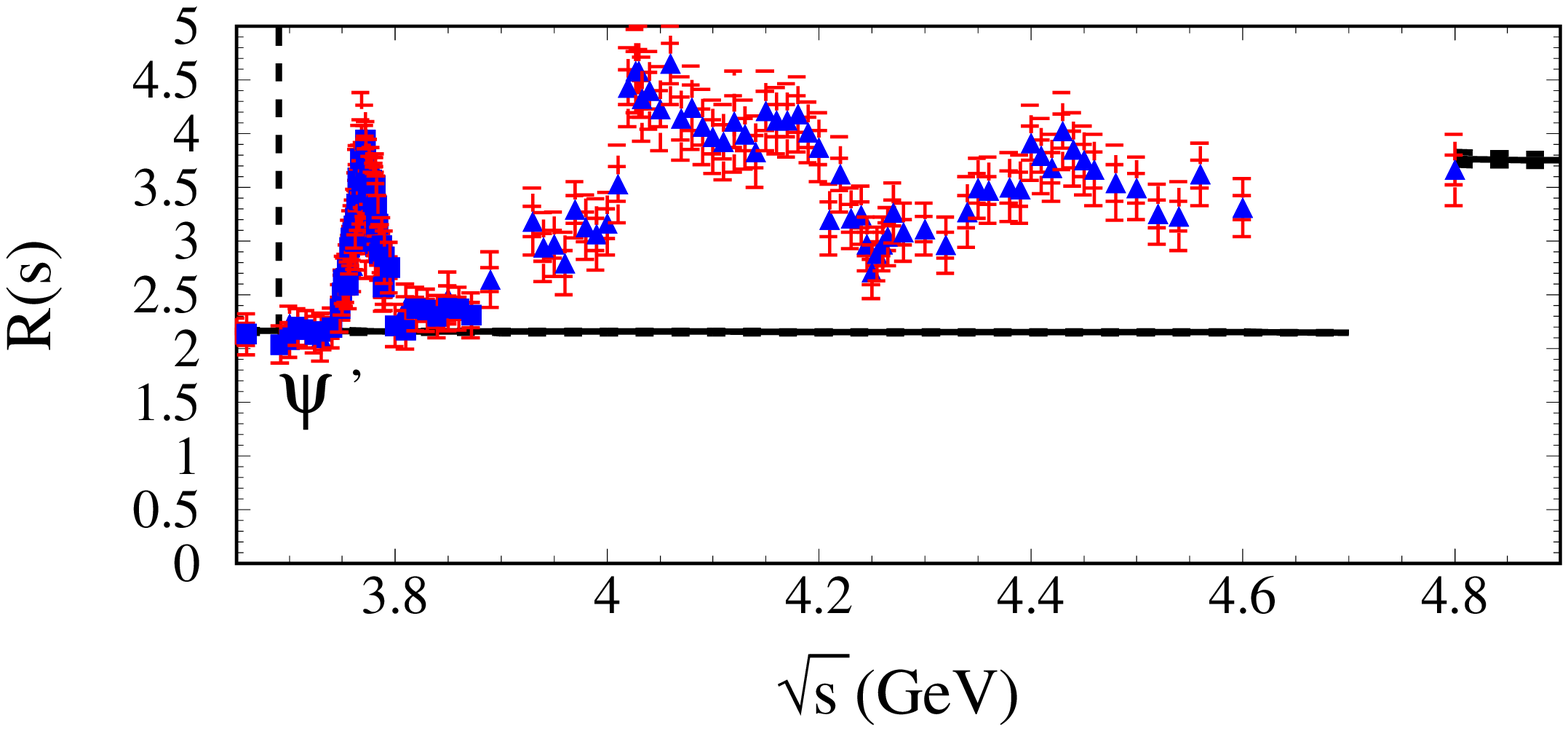}
      \\[-2em]
      \leavevmode
      \epsfxsize=.8\textwidth
      \epsffile{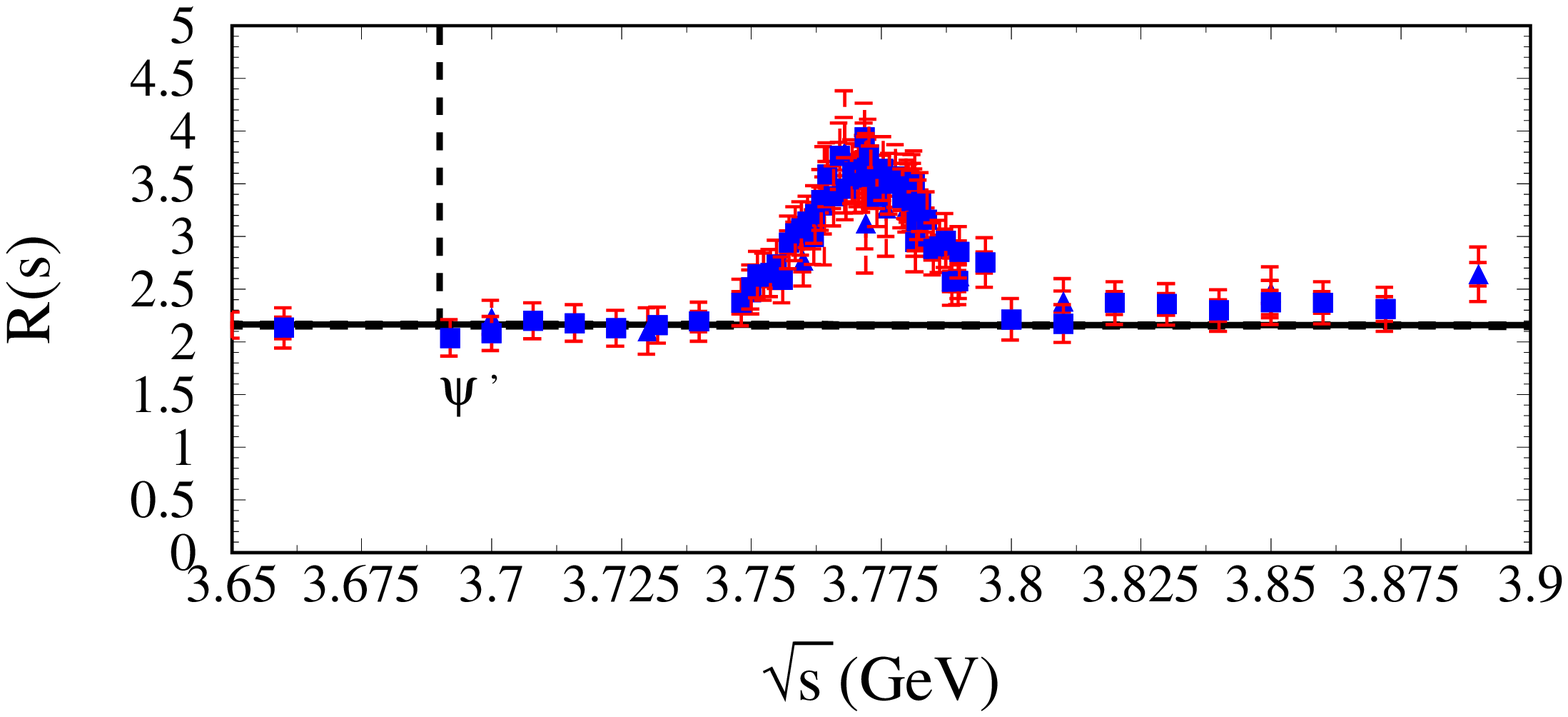}
      \\[-2em]
    \end{tabular}
  \end{center}
  \caption{\label{fig::R}$R(s)$ for different energy intervals around
    the charm threshold region. 
    The solid line corresponds to the theoretical prediction, the  
    uncertainties obtained from the variation of the input parameters
    and of $\mu$  are indicated by the dashed curves.
    The inner and outer error bars give the statistical
    and systematical uncertainty, respectively.
          }
\end{figure}

For the present 
analysis\footnote{We limit this analysis to the results
  from BES~\cite{Bai:2001ct,Ablikim:2006mb}, 
  those from MD-1~\cite{Blinov:1993fw} and from
  CLEO~\cite{Ammar:1997sk} 
  with systematic errors of typically $4.3$\% (BES~\cite{Bai:2001ct}),
  $4$\% (MD-1~\cite{Blinov:1993fw}) and $2$\%  (CLEO~\cite{Ammar:1997sk}). 
  Older measurements, in particular those from
  SPEAR and DORIS, are consistent with the new results. However, with their
  significantly larger errors they do not provide additional
  information.}
we use the data from the BES collaboration
published in 2001~\cite{Bai:2001ct} and
2006~\cite{Ablikim:2006mb}. Whereas the older data cover the whole
range from 2.0~GeV to 4.8~GeV the newer ones provide a very precise 
measurement in the region around $\sqrt{s}\approx 3770$~GeV.
In Fig.~\ref{fig::R} the BES-results are shown together with
the measurements from the MD-1~\cite{Blinov:1993fw} and
CLEO~\cite{Ammar:1997sk} experiments between $4.8$~GeV and $10.52$~GeV.
As is evident from Fig.~\ref{fig::R}, pQCD with
$\alpha_s^{(5)}(M_Z)=0.1189$ provides an excellent description of all
recent results.
For example, the recent BES-value of 
$R^{\rm exp}=2.141\pm0.025({\rm stat.})\pm0.085({\rm syst.})$,
determined in the interval $[3.650\mbox{GeV}, 3.732\mbox{GeV}]$ 
is in excellent agreement with
the theoretical prediction of 
$R(3.700~\mbox{GeV}) = 2.161 \pm  0.017$
and the same is true for
$R^{\rm exp}(10.52~\mbox{GeV}) = 3.56\pm0.01\pm0.07$ 
from CLEO~\cite{Ammar:1997sk} to be compared with
$R(10.52~\mbox{GeV}) = 3.548\pm0.012$.

Up to ${\cal O}(\alpha_s^2)$, the
prediction for charm quark production incorporates the full
quark mass dependence. Starting from order $\alpha_s^2$ also the charm
quark mass dependence of ``secondary'' charm production has to be taken into
account. This includes diagrams of the type in Fig.~\ref{fig::gacc}(a) 
as well as those from Fig.~\ref{fig::gacc}(b).
In addition we include ${\cal O}(\alpha_s^3)$ terms from the expansion
in $(M_c^2/s)^n$ with $n=0,1$ and $2$~\cite{Chetyrkin:1996ia,Chetyrkin:2000zk}.
Last but not least contributions from virtual $c$ quarks
($\sqrt{s}\le3.73$~GeV) and $b$ quarks ($\sqrt{s}\le10.52$~GeV) 
are included. Since our formulae are expressed in terms of
$\alpha_s^{(4)}$, the latter are suppressed $\sim (\alpha_s/\pi)^2
s/(4M_b^2)$ and decouple for $s\ll4M_b^2$.
These are included in the 
forth column of Tab.~\ref{tab::R1}.
Pure QED final state radiation is tiny and taken into account for
completeness.
(For a related analysis at $\sqrt{s} =10.5$~GeV 
see~\cite{Chetyrkin:1996tz}.)

Let us now discuss in detail the evaluation of the ``background''.
The results will be used to  extrapolate $R_{uds}$ from the region
below charm threshold up to 4.8~GeV. 
Here we provide all formulae relevant for the charm
threshold region; the modifications to the bottom case are obvious.

Below $3.73$~GeV only $u$, $d$ and $s$ quarks are produced.
To allow for a smooth transition through the threshold 
the effective number of flavours will be chosen
to be $n_f=4$ and virtual charm quark effects are taken into account
(for a compilation of the relevant formulae
see Refs.~\cite{Chetyrkin:1996ia,Chetyrkin:1997pn,Harlander:2002ur}).

Specifically we define 
\begin{eqnarray}
  R_{\rm background} &=& R_{\rm uds} + R_{\rm uds(cb)} 
  + R_{\rm sing}
  + R_{\rm QED}
  \,,
  \label{eq::Rbackground}
\end{eqnarray}
where we take for the purely light degrees of freedom
\begin{eqnarray}
  R_{\rm uds} &=& 3 \sum_{i=u,d,s} Q_i^2
  \Bigg[1 
  + \frac{\alpha_s}{\pi} 
  + \left(1.640 - 2.250 \lnsmu\right)\left(\frac{\alpha_s}{\pi}\right)^2
  \nonumber\\&&\mbox{}
  + \left(-10.28 - 11.38 \lnsmu + 5.063 \lnsmu^2\right)
  \left(\frac{\alpha_s}{\pi}\right)^3
  \Bigg]
  \,,
  \label{eq::Ruds}
\end{eqnarray}
with $\lnsmu=\ln(s/\mu^2)$ and $\alpha_s\equiv\alpha_s^{(4)}(\mu)$.
Contributions with virtual or secondary charmed quarks appear first in
order $\alpha_s^2$ and in this order they are known in closed analytical
form~\cite{Hoang:1994it,Hoang:1995ex,Chetyrkin:1996yp}:
\begin{eqnarray}
  R_{\rm uds(cb)} &=& 
  \left(\frac{\alpha_s}{\pi}\right)^2
  \left( \delta R_{\rm uds(c)}^{(2)} + \delta R_{\rm uds(b)}^{(2)}
  \right)
  + \left(\frac{\alpha_s}{\pi}\right)^3 \delta R_{\rm uds(c)}^{(3)}
  + \ldots
  \,,
  \nonumber\\
  \delta R_{\rm uds(c)}^{(2)} &=& 
  3 \sum_{i=u,d,s} Q_i^2\,\,
  \frac{2}{3}\left(
  \rho^V(M_c^2,s)+\frac{1}{4}\ln\frac{M_c^2}{\mu^2}
  + \Theta(s-4M_c^2) \rho^R(M_c^2,s)
  \right)
  \,,
  \nonumber\\
  \delta R_{\rm uds(b)}^{(2)} &=& 
  3 \sum_{i=u,d,s} Q_i^2\,\,
  \frac{2}{3} \rho^V(M_b^2,s)
  \label{eq::Rudsc}
  \,,
\end{eqnarray}
where $\rho^V$ and $\rho^R$ can be found in Eqs.~(E.15) and~(E.16) of
Ref.~\cite{Harlander:2002ur}.\footnote{Note that in 
  Ref.~\cite{Harlander:2002ur} $\rho^V$ and
  $\rho^R$ are combined such that $n_f=3$ for $s<4 M_c^2$ and 
  $n_f=4$ for $s>4 M_c^2$ whereas we always have $n_f=4$.}
The first three expansion terms for small and large center-of-mass
energy are given by
\begin{eqnarray}
  \delta R_{\rm uds(c)}^{(2)}\Bigg|_{s\ll 4 M_c^2} &=& 
  3 \sum_{i=u,d,s} Q_i^2 \Bigg[
    \frac{1}{6} \left(\lnsmu-\lnsMc\right)
    + \frac{s}{M_c^2}\left( 0.06519 - 0.01481 \lnsMc \right)
    \nonumber\\&&\mbox{}
    + \left(\frac{s}{M_c^2}\right)^2
    \left( -0.001231 + 0.0003968  \lnsMc \right)
    + \ldots
    \Bigg]
  \,,
  \label{eq::deltaRudsc2_low}
  \\
  \delta R_{\rm uds(c)}^{(2)}\Bigg|_{s\gg 4 M_c^2} &=& 
  3 \sum_{i=u,d,s} Q_i^2 \Bigg[
    -0.1153 + \frac{1}{6} \lnsmu
    + \left(\frac{M_c^2}{s}\right)^2
    \left( -0.4749 + \lnsMc \right)
    + \ldots
    \Bigg]
  \nonumber\\
  \,.
  \label{eq::deltaRudsc2_high}
\end{eqnarray}
$\delta R_{\rm uds(c)}$ exhibits a logarithmic divergence for small
$s$, which could be removed by matching to a theory with three effective
flavours. However, restricting ourselves to $\sqrt{s}>2$~GeV
this logarithm is numerically not large and we can restrict ourselves
to $n_f=4$.
The corresponding $\alpha_s^3$-terms with one or two internal heavy
quark loops, which in Eq.~(\ref{eq::Rudsc}) is parameterized via
$\delta R_{\rm uds(c)}^{(3)}$ are only known in approximative form in
the regions $s\ll 4 M_c^2$~\cite{Larin:1994va,Schreck:dipl} and 
$s\gg 4 M_c^2$~\cite{Chetyrkin:2000zk}:\footnote{The corresponding
  result can easily be extracted from Eqs.~(E.22),~(E.25) and (E.26)
  of Ref.~\cite{Harlander:2002ur}.}
\begin{eqnarray}
  \delta R_{\rm uds(c)}^{(3)}\Bigg|_{s\ll 4 M_c^2} &=& 
  3 \sum_{i=u,d,s} Q_i^2 \Bigg[
    0.1528 
    + 1.005 \lnsmu 
    -  0.7222 \lnsmu^2
    + 0.6944 \lnsmu\lnsMc
    \nonumber\\&&\mbox{}
    - 1.005\lnsMc
    + 0.02778\lnsMc^2
    + \frac{s}{M_c^2}\left( 
    -0.1568
    -0.1709 \lnsmu 
    \right.\nonumber\\&&\mbox{}\left.
    + 0.03210 \lnsmu \lnsMc 
    + 0.1996\lnsMc 
    - 0.03724 \lnsMc^2
    \right)
    \nonumber\\&&\mbox{}
    + \left(\frac{s}{M_c^2}\right)^2
    \left( 
    -0.007178
    + 0.0009988 \lnsmu 
    - 0.00006614 \lnsmu \lnsMc 
    \right.\nonumber\\&&\mbox{}\left.
    - 0.0002714 \lnsMc 
    + 0.0003830 \lnsMc^2
    \right)
    + \ldots
    \Bigg]
  \,,
  \label{eq::deltaRudsc3_low}
  \\
  \delta R_{\rm uds(c)}^{(3)}\Bigg|_{s\gg 4 M_c^2} &=& 
  3 \sum_{i=u,d,s} Q_i^2 \Bigg[
    -1.236 +  1.819 \lnsmu - 0.7222 \lnsmu^2 
    + \frac{M_c^2}{s}\left( - 6.476 \right)
    \nonumber\\&&\mbox{}
    + \left(\frac{M_c^2}{s}\right)^2
    \left( 10.58 + 1.979 \lnsmu + 17.91 \lnsMc 
    -2 \lnsMc^2 
    \right.\nonumber\\&&\left.\mbox{}
    - 4.167 \lnsmu \lnsMc 
    \right)
    + \ldots
    \Bigg]
  \,,
  \label{eq::deltaRudsc3_high}
\end{eqnarray}
with $\lnsMc=\ln(s/M_c^2)$.

\begin{figure}[t]
  \begin{center}
    \begin{tabular}{c}
      \leavevmode
      \epsfxsize=0.95\textwidth
      \epsffile{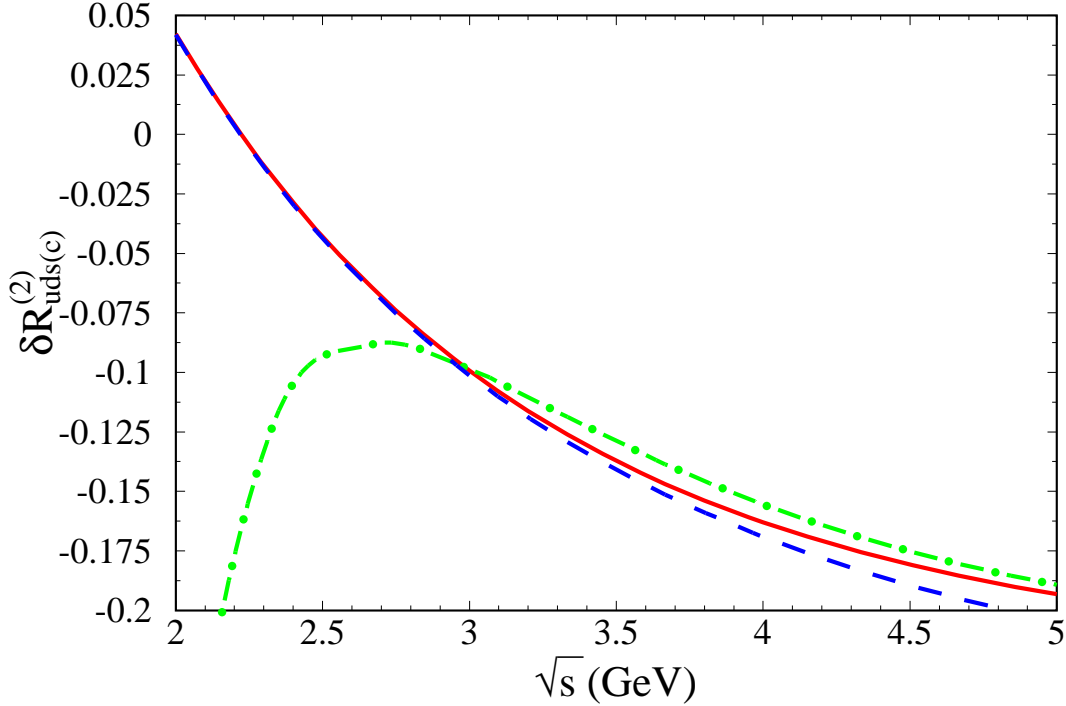}
    \end{tabular}
  \end{center}
  \caption{\label{fig::rudsc}
    $\delta R_{\rm uds(c)}^{(2)}$ for 
    $2~\mbox{GeV}\le \sqrt{s}\le 5~\mbox{GeV}$. Both the exact result
    (solid line) and the approximations including terms of order 
    $(s/M_c^2)^2$ (dashed) and $(M_c^2/s)^2$ (dash-dotted), 
    respectively, are shown and $\mu^2=s$ has been chosen.
    }
\end{figure}

\begin{figure}[t]
  \begin{center}
    \begin{tabular}{c}
      \leavevmode
      \epsfxsize=0.95\textwidth
      \epsffile{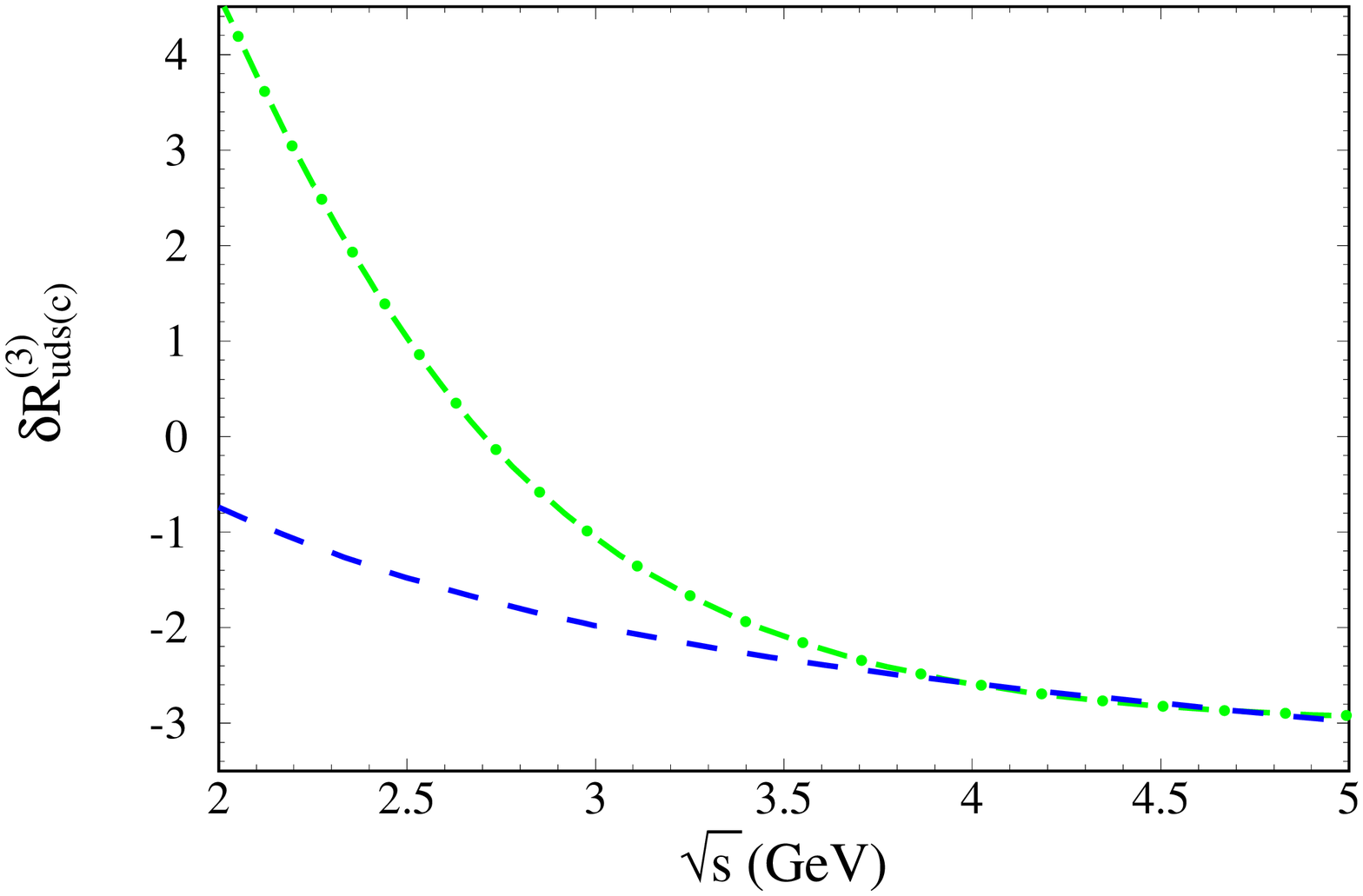}
    \end{tabular}
  \end{center}
  \caption{\label{fig::rudsc3}
    $\delta R_{\rm uds(c)}^{(3)}$ for 
    $2~\mbox{GeV}\le \sqrt{s}\le 5~\mbox{GeV}$. 
    The approximations including terms of order 
    $(s/M_c^2)^2$ (dashed) and $(M_c^2/s)^2$ (dash-dotted), 
    respectively, are shown and $\mu^2=s$ has been chosen.
          }
\end{figure}

In Fig.~\ref{fig::rudsc} the exact result for 
$\delta R_{\rm uds(c)}^{(2)}$ is compared with the approximations of
Eqs.~(\ref{eq::deltaRudsc2_low}) and~(\ref{eq::deltaRudsc2_high}).
As can be seen, the approximations
work excellent even fairly close to threshold, above and
below. 
The corresponding plot for $\delta R_{\rm uds(c)}^{(3)}$ is shown in
Fig.~\ref{fig::rudsc3}. The dash-dotted curve corresponds to 
the approximation in Eq.~(\ref{eq::deltaRudsc3_high})
and the dashed one to Eq.~(\ref{eq::deltaRudsc3_low}). 
One can see a good agreement of the two approximations for
$\sqrt{s} \grtsim 3.5$~GeV. Since for $\sqrt{s} \lessim 3.5$~GeV
the lower dashed curve should be very close to the exact result
we adopt the result in Eq.~(\ref{eq::deltaRudsc3_low}) 
for the whole interval $2~\mbox{GeV} \lessim \sqrt{s} \lessim 5~\mbox{GeV}$.
It is interesting to note that the numerical effect of 
$\delta R_{\rm uds(c)}^{(3)}$ is comparable to the one of
$\delta R_{\rm uds(c)}^{(2)}$. However, their overall size is small,
as can be seen in Tabs.~\ref{tab::R1} and~\ref{tab::R2} (see below).

The singlet contributions start at order $\alpha_s^3$ and are 
fairly small. In the limit $s\gg 4M_c^2$  
$R_{\rm sing}$ can be cast into the form~\cite{Harlander:2002ur}
\begin{eqnarray}
  R_{\rm sing}\Bigg|_{s\gg 4M_c^2} &=& 3 
  \left(\frac{\alpha_s}{\pi}\right)^3
  \left[
    \left(\sum_{i=u,d,s,c} Q_i\right)^2
    \left(-0.4132\right)
    + Q_c \sum_{i=u,d,s,c} Q_i 
    \left(\frac{M_c^2}{s}\right)^2
    17.81
    \right]
  \,.
  \nonumber\\
  \label{eq::Rsing_high}
\end{eqnarray}
There are no quadratic mass terms in the singlet contribution.
Both terms in Eq.~(\ref{eq::Rsing_high}) are proportional to $Q_c^2$,
since the sum over the light quark charges happens to vanish.

The expansion for small $\sqrt{s}$ has been considered in
Refs.~\cite{Larin:1994va,Schreck:dipl} in the context of the $Z$ boson decay.
It is straightforward to adopt Eq.~(20) of~\cite{Larin:1994va} to
our situation for $R_{\rm sing}$. It turns out that 
the lowest four expansion terms for $s\ll 4M_c^2$
are proportional to $Q_u+Q_d+Q_s=0$. Thus the first non-zero contribution is
proportional to $Q_c^2$ and arises at order $(s/M_c^2)^4$. This
term can be found in Refs.~\cite{Groote:2001py,Schreck:dipl}.
However, its contribution is numerically very small and can safely be
neglected in our context.

Combining the information about the small and large energy region for
$R_{\rm sing}$ one obtains that its numerical contribution is between
$-0.55(\alpha_s/\pi)^3$ and zero and can thus be safely neglected.
Similar arguments also hold for the bottom case.

The magnitude of the most important terms contributing to 
$R_{\rm background}$ is shown for several
characteristic energies in Tabs.~\ref{tab::R1} and~\ref{tab::R2} 
for both the charm and the bottom region.
In Tab.~\ref{tab::Rexp} they are compared to a few 
selected experimental results.
Note that in contrast to $R_{\rm background}$
a three-(four-)flavour theory has been used for the $\sqrt{s}$ values
below (above) the charm threshold region.
All theory predictions are given for the reference values
specified in Eq.~(\ref{eq::input}).

\begin{table}[t]
\begin{center}
\begin{tabular}{r|r|r|r|r}
  \hline
  $\sqrt{s}$ (GeV)          & $\alpha_s^{(4)}(\sqrt{s})$ & 
  $R_{\rm uds}$ & $R_{\rm uds(cb)}$  & $R_{\rm background}$ 
  \\
  \hline
$     2.00 $&$   0.3075 $&$   2.2079 $&$   0.0000 $&$   2.2090 $   \\
$     2.30 $&$   0.2882 $&$   2.1952 $&$  -0.0007 $&$   2.1956 $   \\
$     2.60 $&$   0.2733 $&$   2.1853 $&$  -0.0011 $&$   2.1853 $   \\
$     2.90 $&$   0.2614 $&$   2.1773 $&$  -0.0013 $&$   2.1771 $   \\
$     3.20 $&$   0.2516 $&$   2.1706 $&$  -0.0014 $&$   2.1704 $   \\
$     3.50 $&$   0.2433 $&$   2.1650 $&$  -0.0014 $&$   2.1647 $   \\
$     3.80 $&$   0.2362 $&$   2.1601 $&$  -0.0014 $&$   2.1599 $   \\
$     4.10 $&$   0.2300 $&$   2.1559 $&$  -0.0014 $&$   2.1557 $   \\
$     4.40 $&$   0.2245 $&$   2.1522 $&$  -0.0013 $&$   2.1520 $   \\
$     4.70 $&$   0.2197 $&$   2.1489 $&$  -0.0013 $&$   2.1487 $   \\
$     5.00 $&$   0.2153 $&$   2.1459 $&$  -0.0012 $&$   2.1458 $   \\
  \hline
\end{tabular}
\caption{\label{tab::R1}Predictions for $R_{\rm background}$ in
  the charm region for selected energies, separated
  into various pieces as defined in the text.
  For completeness we also give the values of $\alpha_s^{(4)}(\sqrt{s})$.
}
\end{center}
\end{table}

\begin{table}[t]
\begin{center}
\begin{tabular}{r|r|r|r|r}
  \hline
  $\sqrt{s}$ (GeV)          & $\alpha_s^{(5)}(\sqrt{s})$ & 
  $R_{\rm udsc}$ & $R_{\rm udsc(b)}$  & $R_{\rm background}$ 
  \\
  \hline
$     4.80 $&$   0.2186 $&$   3.7617 $&$   0.0010 $&$   3.7649 $   \\
$     5.50 $&$   0.2102 $&$   3.6828 $&$   0.0003 $&$   3.6852 $   \\
$     6.20 $&$   0.2034 $&$   3.6378 $&$  -0.0003 $&$   3.6396 $   \\
$     6.90 $&$   0.1977 $&$   3.6092 $&$  -0.0006 $&$   3.6107 $   \\
$     7.60 $&$   0.1928 $&$   3.5896 $&$  -0.0009 $&$   3.5908 $   \\
$     8.30 $&$   0.1886 $&$   3.5753 $&$  -0.0011 $&$   3.5764 $   \\
$     9.00 $&$   0.1849 $&$   3.5644 $&$  -0.0012 $&$   3.5654 $   \\
$     9.70 $&$   0.1816 $&$   3.5559 $&$  -0.0013 $&$   3.5567 $   \\
$    10.40 $&$   0.1786 $&$   3.5489 $&$  -0.0014 $&$   3.5497 $   \\
$    11.10 $&$   0.1759 $&$   3.5431 $&$  -0.0014 $&$   3.5438 $   \\
$    11.80 $&$   0.1735 $&$   3.5382 $&$  -0.0014 $&$   3.5389 $   \\
  \hline
\end{tabular}
\caption{\label{tab::R2}Predictions for $R_{\rm background}$ in
  the bottom region for selected energies, separated
  into various pieces as defined in the text.
  For completeness we also give the values of $\alpha_s^{(5)}(\sqrt{s})$.
}
\end{center}
\end{table}

\begin{table}[t]
{\scalefont{0.8}
\begin{center}
\begin{tabular}{c|llllll}
  \hline
  $\sqrt{s}$ (GeV) & 2.00      & 3.65      & 3.732 
                   & 4.80      & 9.00      & 10.52 
  \\
  $R^{\rm th}(s)$  & 2.209(91) & 2.161(18) & 2.160(17) 
                   & 3.764(64) & 3.564(17) & 3.548(12)
  \\
  $R^{\rm exp}(s)$ & 2.18(7)(18) & 2.157(35)(86) & 2.156(86)(86)
                   & 3.66(14)(19)& 3.62(7)(14)   & 3.56(1)(7)
  \\
  Experiment & 
  BES~\cite{Bai:2001ct} & BES~\cite{Ablikim:2006mb} & BES~\cite{Ablikim:2006mb}
  &
  BES~\cite{Bai:2001ct} & MD-1~\cite{Blinov:1993fw}  & CLEO~\cite{Ammar:1997sk}
  \\
  \hline
\end{tabular}
\caption{\label{tab::Rexp}Comparison of the theory predictions for 
  $R(s)$ with the experimental results at a few selected 
  values for $\sqrt{s}$.
}
\end{center}
}
\end{table}

The band in Fig.~\ref{fig::R} and the uncertainty in
Tab.~\ref{tab::Rexp} is obtained from the parametric
errors and the theoretical uncertainty, the latter obtained by varying the
renormalization scale between $\mu^2=s/4$ and $\mu^2=4s$. These were
added quadratically.
The excellent agreement between prediction and measurement 
in the continuum justifies to use the background formula, normalized to
data below threshold also in the region above to obtain
as remainder $R_c$ which will be used for the quark mass determination.

In passing let us mention that the formulae in combination with the
$R$ measurement between $3$~GeV and $10.5$~GeV can be used to
determine a fairly precise value for $\alpha_s$, as demonstrated in
Ref.~\cite{Kuhn:2001dm}. 


\section{\label{sec::charm}Charm 
  quark mass determination from the threshold region}

Our determination of the charmed quark mass follows closely
Ref.~\cite{Kuhn:2001dm}. It is based on the direct comparison of the
theoretical and experimental evaluations of the contributions to the
derivatives of the polarization function $\Pi_c(q^2)$ (as defined in
Eq.~(\ref{eq::pivadef})), the former evaluated in perturbative QCD,
the latter through moments of the measured cross section for charm
production in electron-positron annihilation. In its domain of
analyticity $\Pi_c(q^2)$ can be cast into the form
\begin{eqnarray}
  \Pi_c(q^2) &=& Q_c^2 \frac{3}{16\pi^2} \sum_{n\ge0}
                       \bar{C}_n z^n
  \,,
  \label{eq::pimom}
\end{eqnarray}
with $Q_c=2/3$ and 
$z=q^2/(4m_c^2)$ where $m_c=m_c(\mu)$ is the $\overline{\rm MS}$ 
charm quark mass at the scale $\mu$.
The perturbative series for the coefficients $\bar{C}_n$ 
in order $\alpha_s^2$ was evaluated up to $n=8$ in
Ref.~\cite{Chetyrkin:1995ii,Chetyrkin:1997mb} 
(recently~\cite{Boughezal:2006uu} even up to $n=30$). The four-loop contributions
to $\bar{C}_0$ and $\bar{C}_1$ were calculated in
Ref.~\cite{Chetyrkin:2006xg,Boughezal:2006px}.
The coefficients depend on $\alpha_s$ and on the charm quark mass
through logarithms of the form $\lmc\equiv\ln(m_c^2(\mu)/\mu^2)$. They
can be written as 
\begin{eqnarray}
  \bar{C}_n &=& \bar{C}_n^{(0)} 
  + \frac{\alpha_s(\mu)}{\pi}
  \left( \bar{C}_n^{(10)} + \bar{C}_n^{(11)}\lmc \right)
  + \left(\frac{\alpha_s(\mu)}{\pi}\right)^2
  \left( \bar{C}_n^{(20)} + \bar{C}_n^{(21)}\lmc
  + \bar{C}_n^{(22)}\lmc^2 \right)
  \nonumber\\&&\mbox{}
  + \left(\frac{\alpha_s(\mu)}{\pi}\right)^3
  \left( \bar{C}_n^{(30)} + \bar{C}_n^{(31)}\lmc
  + \bar{C}_n^{(32)}\lmc^2 + \bar{C}_n^{(33)}\lmc^3 \right)
  + \ldots
  \,.
  \label{eq::cn}
\end{eqnarray}
In Tab.~\ref{tab::cn} the coefficients $\bar{C}_n^{(ij)}$ are given 
in numerical form up to $n=4$.
The terms $C_n^{(30)}$ with $n\ge2$ are still unknown.
To estimate the theoretical uncertainly from this source we will use the
following bounds on $C^{(30)}_n$:
\begin{eqnarray}
  \label{eq::growth}
  -6.0 \le \bar{C}_2^{(30)} \le 7.0\,,\qquad
  -6.0 \le \bar{C}_3^{(30)} \le 5.2\,,\qquad
  -6.0 \le \bar{C}_4^{(30)} \le 3.1\,.
\end{eqnarray}
The lower limit $C_n^{(30)} > -6.0$ is based
on the observation that for fixed order in $\alpha_s$ the absolute
values of $C_n^{(20)}$, $C_n^{(10)}$ and $C_n^{(0)}$ in only one case
exhibit a slight increase, in general, however, they tend to decrease
with increasing $n\ge1$ quite markedly. The upper limit is estimated
from the dependence on the order in $\alpha_s$ with $n$ fixed, where
we assume at maximum a geometric increase and thus $C_n^{(30)} \le
(C_n^{(20)})^2/C_n^{(10)}$. 
In our numerical analysis we also include the two-loop QED corrections 
to $C_n$ which can easily be obtained from $C_n^{(20)}$. It leads to a
negative shift in $m_c$ by less than $2$~MeV.

Tab.~\ref{tab::cn} and Eq.~(\ref{eq::growth})
essentially constitutes our theoretical input.
We define the moments
\begin{eqnarray}
  {\cal M}_n &\equiv& \frac{12\pi^2}{n!}
                      \left(\frac{{\rm d}}{{\rm d}q^2}\right)^n
                      \Pi_c(q^2)\Bigg|_{q^2=0}
  \,,
\end{eqnarray}
which leads to
\begin{eqnarray}
  {\cal M}_n^{\rm th} &=& 
  \frac{9}{4}Q_c^2
  \left(\frac{1}{4 m_c^2}\right)^n \bar{C}_n
  \,.
  \label{eq::Mth}
\end{eqnarray}
As demonstrated in Ref.~\cite{Kuhn:2001dm}, high moments ($n>4$) are less
suited for the mass extraction. The analysis will therefore be
limited to $n<4$ and the results for $n=4$ will only be presented for illustration.

\renewcommand{\arraystretch}{1.1}
\begin{table}
\begin{center}
{\footnotesize
\begin{tabular}{r|rrrrrrrrrr}
\hline
  $n$ & 
  $\bar{C}_n^{(0)}$  & $\bar{C}_n^{(10)}$ & $\bar{C}_n^{(11)}$ & 
  $\bar{C}_n^{(20)}$ & $\bar{C}_n^{(21)}$ & $\bar{C}_n^{(22)}$ & 
  $\bar{C}_n^{(30)}$ & $\bar{C}_n^{(31)}$ & $\bar{C}_n^{(32)}$ & 
  $\bar{C}_n^{(33)}$
  \\
  \hline
1&$   1.0667$&$   2.5547$&$   2.1333$&$   2.4967$&$   3.3130$&$  -0.0889$&$  -5.6404$&$   4.0669$&$   0.9590$&$   0.0642$\\
2&$   0.4571$&$   1.1096$&$   1.8286$&$   2.7770$&$   5.1489$&$   1.7524$&$$---$$&$   6.7216$&$   6.4916$&$  -0.0974$\\
3&$   0.2709$&$   0.5194$&$   1.6254$&$   1.6388$&$   4.7207$&$   3.1831$&$$---$$&$   7.5736$&$  13.1654$&$   1.9452$\\
4&$   0.1847$&$   0.2031$&$   1.4776$&$   0.7956$&$   3.6440$&$   4.3713$&$$---$$&$   4.9487$&$  17.4612$&$   5.5856$\\
\hline
\end{tabular}
}
\caption{\label{tab::cn}Coefficients of the photon polarization
  function in the $\overline{\rm MS}$ scheme as defined in
  Eqs.~(\ref{eq::pimom}) and~(\ref{eq::cn}).
  $n_f=4$ has been adopted which is appropriate for the charm threshold.
  }
\end{center}
\end{table}
\renewcommand{\arraystretch}{1.0}

With the help of the dispersion relation we establish the connection
between the polarization function and the experimentally accessible 
cross section $R_c(s)$. In the $\overline{\rm MS}$ scheme
\begin{eqnarray}
  \Pi_c(q^2) &=& \frac{q^2}{12\pi^2}\int
  {\rm d}s\,\frac{R_c(s)}{s(s-q^2)}
  + Q_c^2 \frac{3}{16\pi^2} \bar{C}_0 
  \,,
  \label{eq::pidisp}
\end{eqnarray}
which allows to determine the experimental moments
in Eq.~(\ref{eq::Mexp}).
Note, that the last term in Eq.~(\ref{eq::pidisp}) 
which defines the renormalization scheme disappears after
taking derivatives with respect to $q^2$.
Equating Eqs.~(\ref{eq::Mth}) and~(\ref{eq::Mexp}) 
leads to an expression from which the charm quark mass can be obtained:
\begin{eqnarray}
  m_c(\mu) &=& \frac{1}{2} 
  \left(\frac{\bar{C}_n}{{\cal M}_n^{\rm exp}}\right)^{1/(2n)} 
  \,.
  \label{eq::mc1}
\end{eqnarray}

Non-perturbative contributions to the moments have been evaluated
in~\cite{Novikov:1977dq,Broadhurst:1994qj} and, in view of their
smallness\footnote{This observation was confirmed in Ref.~\cite{Hoang:2004xm}.},
were neglected in Ref.~\cite{Kuhn:2001dm}.
With the further reduction of theoretical and experimental
uncertainties this approximation has to be reconsidered.

The non-perturbative contribution from gluon
condensate is given by~\cite{Novikov:1977dq,Broadhurst:1994qj}
\begin{eqnarray}
  \delta {\cal M}^{\rm np}_n &=&
  \frac{12\pi^2 Q_c^2}{(4 m_c^2)^{(n+2)}}
  \left\langle\frac{\alpha_s}{\pi} G^2 \right\rangle
  \,a_n\,\left(1+\frac{\alpha_s}{\pi}\bar{b}_n\right)
  \,,
  \label{eq::Mnp}
\end{eqnarray}
with
\begin{eqnarray}
  a_n &=& - \frac{2n+2}{15} 
  \frac{\Gamma(4+n)\Gamma\left(\frac{7}{2}\right)}
  {\Gamma(4)\Gamma\left(\frac{7}{2}+n\right)}
  \,,\nonumber\\
  \bar{b}_n &=& b_n - \left( 2n+4 \right)
  \left(\frac{4}{3}-\lmc\right)
  \,,
\end{eqnarray}
and $b_n$ taken from Tab.~2 of Ref.~\cite{Broadhurst:1994qj}:
\begin{eqnarray}
  b_1\,\,=\,\,\frac{135779}{12960}\,,\quad
  b_2\,\,=\,\,\frac{1969}{168}\,,\quad
  b_3\,\,=\,\,\frac{546421}{42525}\,,\quad
  b_4\,\,=\,\,\frac{661687433}{47628000}\,.
\end{eqnarray}
For the numerical value of the condensate we take~\cite{Ioffe:2005ym}
\begin{eqnarray}
  \left\langle \frac{\alpha_s}{\pi} G^2 \right\rangle 
  &=& 0.006 \pm 0.012~(\mbox{GeV})^4
  \,.
\end{eqnarray}
For the evaluation of $\delta{\cal M}_n^{\rm np}$ we use 
$\alpha_s(3~\mbox{GeV})=0.254$, $\mu=3~\mbox{GeV}$
and $m_c=1.3$~GeV. The result for
$\delta{\cal M}_n^{\rm np}$ are listed in Tab.~\ref{tab::Mexp} and
will be subtracted from $\delta{\cal M}_n^{\rm exp}$
which enters Eq.~(\ref{eq::mc1}).
Their impact on $m_c$ is still negligible,
although the contribution to the error starts to become relevant for $n\ge3$.

Let us now turn to the three different 
contributions which enter the right-hand side of Eq.~(\ref{eq::Mexp}):
the $J/\Psi$ and $\Psi^\prime$
resonances (${\cal M}_n^{\rm res}$), the charm threshold region between 
$2M_{D_0}\approx3.73$~GeV and $\sqrt{s_1}=4.8$~GeV as measured by the BES
experiment~\cite{Bai:2001ct,Ablikim:2006mb}
(${\cal M}_n^{\rm thresh}$), and the continuum contribution above
$s_1$ (${\cal M}_n^{\rm cont}$).

The two narrow resonances are treated in the narrow width approximation
\begin{eqnarray}
  R^{\rm res}(s) &=& \frac{9\pi M_R \Gamma_e}{\alpha^2}
                     \left(\frac{\alpha}{\alpha(s)}\right)^2
                     \delta(s-M_R^2)
  \,,
\end{eqnarray}
where we use the parameters listed in Tab.~\ref{tab::psi}.

\begin{table}[t]
  \begin{center}
    {
      \begin{tabular}{c|c|c}
	\hline
        &$J/\Psi$     &$\Psi(2S)$    \\ \hline
	$M_{\Psi}$(GeV)              & 3.096916(11) & 3.686093(34) \\
	$\Gamma_{ee}$(keV)           & 5.55(14)     & 2.48(6)      \\
	$(\alpha/\alpha(M_{\Psi}))^2$& 0.957785     & 0.95554      \\
	\hline
      \end{tabular}
    }
    \caption{\label{tab::psi} Masses and electronic widths~\cite{Yao:2006px} 
      of the narrow charmonium resonances and effective electromagnetic 
      coupling~\cite{Teubner:priv_com} at the appropriate scales.}
  \end{center}
\end{table}
 
In the charm threshold region (which includes $\Psi(3770)$)
we have to identify the contribution from
the charm quark, i.e. we have to subtract the parts
arising from the light $u$, $d$ and $s$ quark.
(As discussed in Section~\ref{sec::background}, charm production
through secondary gluon splitting is treated as part of the
background, the same applies to the singlet contribution.)
Technically this is done by determining a mean value for 
$R_{\rm background}$ (cf. Eq.~(\ref{eq::Rbackground}))
from the comparison of theoretical
predictions and the BES data between $2$~GeV and
$3.73$~GeV and using the theoretically predicted energy dependence 
to extrapolate into the region between $3.73$~GeV and
$4.8$~GeV~\cite{Kuhn:1998ze}.
The mean value at the energy just below the $D\bar{D}$-threshold is
given by
\begin{eqnarray}
  \bar{R}_{\rm background} &=& R^{\rm th}(s_-) \, n_{-}
  \,,
\end{eqnarray}
with
\begin{eqnarray}
  n_{-} &=&
  \sum_{s_i\le s_-}
  \frac{ R^{\rm exp}(s_i) }{R^{\rm th}(s_i)}
  \frac{\sigma_-^2}
       {\left(\sigma_i\frac{R^{\rm th}(s_-)}{R^{\rm th}(s_i)}  
	 \right)^2}
       \,,\nonumber\\
  \frac{1}{\sigma_-^2} &=&
  \sum_{s_i\le s_-} 
  \frac{1}
       {\left(\sigma_i\frac{R^{\rm th}(s_-)}{R^{\rm th}(s_i)}  
	 \right)^2}
       \,,
\end{eqnarray}
where the sums run over the $s$-values where experimental data,
$R^{\rm exp}$, are available. In our case the upper bound is given by
$\sqrt{s_-}=3.73$~GeV and $R^{\rm th}$ corresponds to the 
theoretical predictions from $R_{\rm background}(s)$.
In a next step we multiply $R_{\rm background}(s)$ with the
normalization factor $n_{-}$ and subtract the result for each
$s$-value in the energy region $[3.73~\mbox{GeV},4.8~\mbox{GeV}]$ from
the data before the integration is performed. The final result for the charm quark mass
$m_c(3~\mbox{GeV})$ changes by about $-5$~MeV 
if we do not take this
renormalization into account.
The effect of the energy dependence of $R_{\rm background}$ 
has practically no effect on the final result (less than 1~MeV).
The normalization factors 
$n_{-} = 1.038$ 
for the 2001 data~\cite{Bai:2001ct} and 
$n_{-} = 0.991$ 
for the 2006 data~\cite{Ablikim:2006mb}
can be considered as systematic offset of the experimental $R$
measurement, as determined from the measurement below threshold and is
determined under the assumption that the energy dependence of the true
$R$-ratio is indeed given by $R^{\rm th}$. The magnitude of this factor
is consistent with the quoted systematic error 
of 4.3\%~\cite{Bai:2001ct} and of 4.0\% (4.9\%) outside
(within) the $\Psi(3770)$-region~\cite{Ablikim:2006mb} 
which is kept in the subsequent analysis.
The statistical error of $n_{-}$ is negligible. 

In the continuum region above $\sqrt{s}=4.8$~GeV data are sparse and
imprecise. On the other hand, pQCD provides reliable predictions for
$R(s)$, which is essentially due to the knowledge of the complete mass
dependence up to order 
$\alpha_s^2$~\cite{Chetyrkin:1995ii} and the dominant terms of order
$\alpha_s^3$~\cite{Chetyrkin:2000zk}.
Thus in this region we will replace data by the theoretical prediction
for $R(s)$ as discussed in the beginning of Section~\ref{sec::background}.
As emphasized above, the cross section is only weakly $m_c$-dependent
in this region.

In Tab.~\ref{tab::Mexp} we present the results for the moments
separated according to the three different contributions
discussed above.
The error of the resonance contribution is due to the uncertainties
of the input parameters. The error of the charm threshold
contribution is dominated by the correlated
normalization error of approximately 4.0\% to 4.9\% of the BES data. 

\begin{table}[t]
\begin{center}
{
\begin{tabular}{l|lll|l||l}
\hline
$n$ & ${\cal M}_n^{\rm res}$
& ${\cal M}_n^{\rm thresh}$
& ${\cal M}_n^{\rm cont}$
& ${\cal M}_n^{\rm exp}$
& ${\cal M}_n^{\rm np}$
\\
& $\times 10^{(n-1)}$
& $\times 10^{(n-1)}$
& $\times 10^{(n-1)}$
& $\times 10^{(n-1)}$
& $\times 10^{(n-1)}$
\\
\hline
$1$&$  0.1201(25)$ &$  0.0318(15)$ &$  0.0646(11)$ &$  0.2166(31)$ &$ -0.0001(2)$ \\
$2$&$  0.1176(25)$ &$  0.0178(8)$ &$  0.0144(3)$ &$  0.1497(27)$ &$ 0.0000(0)$ \\
$3$&$  0.1169(26)$ &$  0.0101(5)$ &$  0.0042(1)$ &$  0.1312(27)$ &$  0.0007(14)$ \\
$4$&$  0.1177(27)$ &$  0.0058(3)$ &$  0.0014(0)$ &$  0.1249(27)$ &$  0.0027(54)$ \\
\hline
\end{tabular}
}
\caption{
  \label{tab::Mexp}Experimental moments 
  in $(\mbox{GeV})^{-2n}$ as defined in
  Eq.~(\ref{eq::Mexp}) separated according to the contributions from
  the narrow resonances,
  the charm threshold region and the continuum region
  above $\sqrt{s}=4.8$~GeV.
  In the last column the contribution from the gluon condensate is shown.}
\end{center}
\end{table}

To estimate the error on ${\cal M}_n^{\rm cont}$ we varied the input
parameters as stated in Eq.~(\ref{eq::input}) and 
the renormalization scale $\mu$ between $\sqrt{s}/2$ and $2\sqrt{s}$.
The errors of the three contributions are added quadratically.
It is instructive to compare the composition
of the experimental error for the different moments. Generally speaking,
it is dominated by the resonance contribution, specifically by the
2.5\% uncertainty in the leptonic widths of the $J/\Psi$ and
$\Psi^\prime$.
Compared to the analysis of~\cite{Kuhn:2001dm} the results are
consistent. However, the present error amounts to less than half of the
previous one.
Already for the previous analysis the improvement in the
cross section measurement due to BES
(from about 10--20 \% systematic error down to 4.3\%) was important. 
This is particularly true for the moment $n=1$.
The recent data for the $\Psi(3770)$-region have lead to further improvement.
The parametric uncertainties (from $\alpha_s$ and $M_c$) and the residual
$\mu$-dependence which affect ${\cal M}_n^{\rm cont}$ are small. The higher
moments 
(in fact already for $n$ above two) are increasingly dominated by
the resonance contributions with their 2.5\% uncertainty. 
The large uncertainty in
$\langle \alpha_s/\pi G^2 \rangle$ starts to matter for $n\ge3$.

\begin{table}[t]
\begin{center}
{
\begin{tabular}{l|l|llll|l|l|l}
\hline
&&&&&&&&\\[-.8em]
$n$ & $m_c(3~\mbox{GeV})$ & 
exp & $\alpha_s$ & $\mu$ & np & 
total & $\delta\bar{C}_n^{(30)}$ &
$m_c(m_c)$ 
\\
\hline
        1&  0.986&  0.009&  0.009&  0.002&  0.001&  0.013&---&  1.286 \\
        2&  0.979&  0.006&  0.014&  0.005&  0.000&  0.016&  0.006&  1.280 \\
        3&  0.982&  0.005&  0.014&  0.007&  0.002&  0.016&  0.010&  1.282 \\
        4&  1.012&  0.003&  0.008&  0.030&  0.007&  0.032&  0.016&  1.309 \\
\hline
\end{tabular}
}
\caption{\label{tab::mc1}Results for $m_c(3~\mbox{GeV})$ in GeV
  obtained from Eq.~(\ref{eq::mc1}). The errors are from experiment,
  $\alpha_s$, variation of $\mu$
  and the condensate.
  The error from the yet unknown four-loop term is kept separate.
  Last column: central values for $m_c(m_c)$.
}
\end{center}
\end{table}

We use the results of Tab.~\ref{tab::Mexp} together with
Eqs.~(\ref{eq::mc1}) and~(\ref{eq::Mnp})
in order to obtain in a first step $m_c(3~\mbox{GeV})$.
The results are listed in Tab.~\ref{tab::mc1} together with the total
uncertainty and its decomposition into the individual contributions
from the experimental moments, the variation of $\alpha_s$
(cf. Eq.~(\ref{eq::input})), the renormalization scale
$\mu=(3\pm1)$~GeV and the condensate which are all added quadratically. 
To determine the $\mu$-dependence, we evaluate Eq.~(\ref{eq::mc1}) at
$\mu=2$~GeV and 4~GeV. Subsequently, using the renormalization group
equation to four-loop accuracy,
we evolve the result for $m_c(\mu)$ to the reference point
$\mu=3$~GeV and compare with the original value. 

The first moment is available in four-loop approximation, and the
corresponding $\mu$-dependence of the result is completely
negligible. It is interesting to anticipate the corresponding
$\mu$-dependence for the four-loop result also for the higher moments.
The corresponding results are displayed in Tab.~\ref{tab::mc1}.
In the eighth column we show the effect of the unknown coefficients
$\bar{C}_n^{(30)}$ ($n\ge2$) where assumptions were made on their
respective numerical growth as discussed above.
For the extraction of $m_c(3~\mbox{GeV})$ we assume
$\bar{C}_n^{(30)}=0$.
For $\delta \bar{C}_n^{(30)}$ we take the larger of the deviations
estimated in Eq.~(\ref{eq::growth}).

Let us compare the results displayed in Tab.~\ref{tab::mc1} with those
from~\cite{Kuhn:2001dm}. For the moment with $n=1$ the main
improvement originates from the new experimental results for the
electronic widths of the $J/\Psi$ and $\Psi^\prime$. 
Some additional improvement is due to the new BES-data~\cite{Ablikim:2006mb}.
The four-loop
term leads to a further reduction of the theoretical uncertainty ---
in view of the dominant experimental uncertainties it only leads to
consolidation of the theoretical prediction.
The error composition in the analysis based on $n=2$ is similar, 
as far as the
experimental input is concerned. However, in this case the error from
the $R$-measurements is somewhat smaller, the error induced by the 
uncertainty in $\alpha_s$ is larger. Furthermore, the error from 
$\delta C^{(30)}_2$ has to be taken into account. Although 
at first glance this
error looks small, a check of the assumptions on
$C^{(30)}_2$ would be extremely useful.

The analysis based on $n=3$ leads to a
result consistent with the one for $n=1$ and $n=2$. In
this case, however, the theoretical uncertainty 
estimated through the $\mu$ dependence starts to become important and
the total error is therefore larger.

For the subtraction of the light-quark continuum we have assumed pQCD
to be exact. Deviations from this idealized
situation have been estimated in Ref.~\cite{Shifman:2003de}.
Based on theoretical and phenomenological considerations it is assumed,
that the oscillations of $R(s)$ around the perturbative result, which
are observed in the low-energy region and reflect the low-lying
resonances, are damped by a power-law or an exponential factor. The
following two correction factors were suggested in~\cite{Shifman:2003de}
\begin{eqnarray}
  f_1 &=& 1 + 1.22 \, s^{-3/2} \,
  \sin\left(2\rho\sqrt{s}-\delta\right)
  \,,
\end{eqnarray}
with $\rho=3~\mbox{GeV}^{-1}$, $\delta=1.32$ and, alternatively,
\begin{eqnarray}
  f_2 &=& 1 - 1.24\exp\left(-\frac{2\pi s B}{\sigma^2 N_c}\right)
  \sin\left(\frac{2\pi s}{\sigma^2}-3.08\right)
  \,,
\end{eqnarray}
with $\sigma^2=2~\mbox{GeV}^2$, $B=0.5$ and $N_c=3$.
Potential effects from this modification are estimated 
by including the factors $f_1$ or $f_2$ into the 
analysis.\footnote{We thank A. Vainshtein for raising this question.}
In the case of $f_1$ we also vary the phase $\delta$ between 0 and $\pi$.

The change in the contribution from the region 
$3.73~\mbox{GeV}\le \sqrt{s}\le 4.8~\mbox{GeV}$ is about 1.2\% for $f_1$
and $n=1$ and significantly smaller in the case of $f_2$. The 1.2\%
translate to a shift of $-1$~MeV for $m_c(3~\mbox{GeV})$
which is completely negligible in the present context.

\begin{figure}[t]
  \begin{center}
    \begin{tabular}{c}
      \leavevmode
      \epsfxsize=0.95\textwidth
      \epsffile{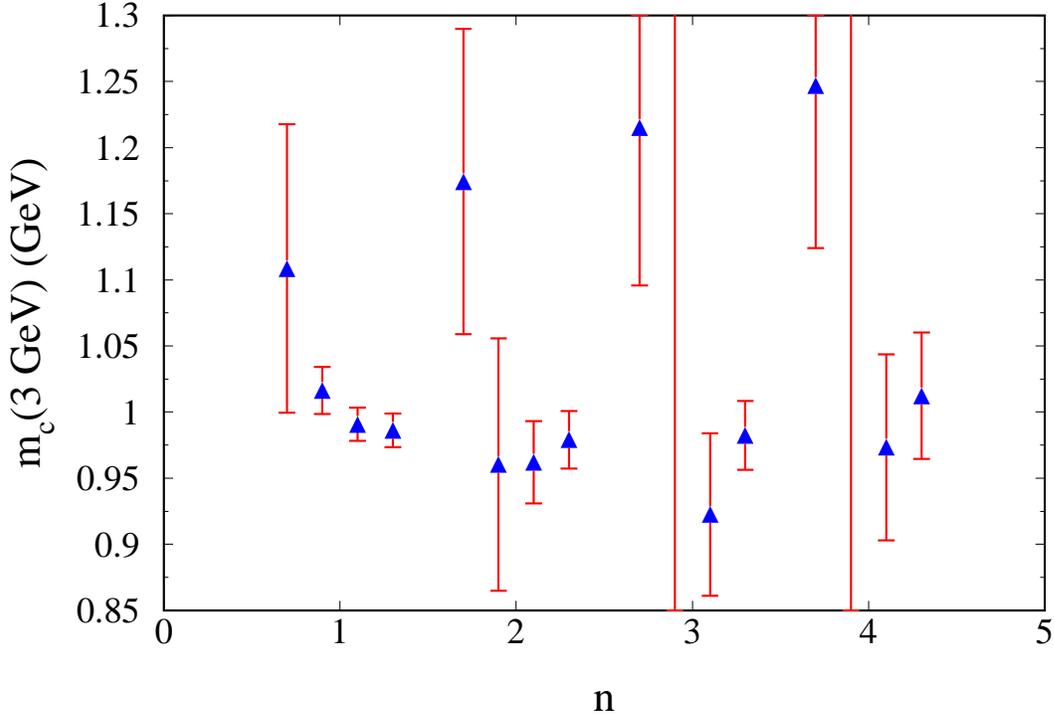}
    \end{tabular}
  \end{center}
  \vspace*{-2em}
  \caption{
    \label{fig::mom}$m_c(3~\mbox{GeV})$ for $n=1,2,3$ and $4$.
    For each value of $n$ the results from left to right correspond
    the inclusion of terms of order $\alpha_s^0$, $\alpha_s^1$,
    $\alpha_s^2$ and $\alpha_s^3$ in the coefficients $\bar{C}_n$ 
    (cf. Eq.~(\ref{eq::cn})).
    Note, that for $n=3$ and $n=4$ the central values and
    uncertainties can not be determined
    with the help of Eq.~(\ref{eq::mc1})
    in those cases where only the two-loop corrections of order $\alpha_s$ are
    included into the coefficients $\bar{C}_n$ as the equation cannot be
    solved for $m_c(3~\mbox{GeV})$.
  }
\end{figure}

The results of Tab.~\ref{tab::mc1} are also shown in
Fig.~\ref{fig::mom} where the central value and the uncertainties of
$m_c(3~\mbox{GeV})$ are plotted for $n=1,2,3$ and $4$. For each $n$
the results are shown for one- to four-loop theory input.
For the ${\cal O}(\alpha_s^3)$ term
the exact result is used for $n=1$ whereas for $n=2,3$
and 4 we use $C_n^{(30)}=0$ and the error estimates discussed above.
It is nicely seen that the results for 
$m_c(3~\mbox{GeV})$ further stabilize when going from three to four
loops. At the same time the uncertainty is considerably reduced.
Furthermore, the preference for the first three moments is 
clearly visible.
Also the analysis for $n=2$ and $n=3$ leads to small errors, even if we
include to contribution from the yet uncalculated $\delta C^{(30)}_n$.
We emphasize the remarkable consistency between the three results which
we consider as additional confirmation of our approach.

The result based on the 
moment with $n=1$ is evidently least sensitive to non-perturbative
contributions from condensates, to the Coulombic higher order effects, the
variation of $\mu$ and the parametric $\alpha_s$ dependence.
The results for $n=2$ and $n=3$ are practically identical. Since the
error in this case is somewhat larger, and furthermore affected by the
estimate for $\delta C_n^{(30)}$, we adopt
\begin{eqnarray}
  m_c(3~\mbox{GeV}) &=& 0.986(13)~\mbox{GeV}
  \,,
  \label{eq::mc3final}
\end{eqnarray}
as our final result. 
Transforming this to the scale-invariant mass
$m_c(m_c)$~\cite{Chetyrkin:2000yt}
including the four-loop coefficients of the renormalization group
functions one finds
\begin{eqnarray}
  m_c(m_c) &=& 1.286(13)~\mbox{GeV}
  \,.
  \label{eq::mcmc}
\end{eqnarray}
Our result agrees within the uncertainties with our last
determination~\cite{Kuhn:2001dm}, but is considerably more precise.
Using the three-loop 
relation~\cite{Chetyrkin:1999ys,Chetyrkin:1999qi,Melnikov:2000qh}
between pole- and $\overline{\rm MS}$-mass this corresponds to
$M_c^{\rm(3-loop)} = 1.666~\mbox{GeV}$.
We refrain from providing an uncertainty for $M_c$ since it is well known
that the perturbative series between the $\overline{\rm MS}$ and the
pole mass is badly behaved.

\begin{figure}[t]
  \begin{center}
    \begin{tabular}{c}
      \leavevmode
      \epsfxsize=0.95\textwidth
      \epsffile{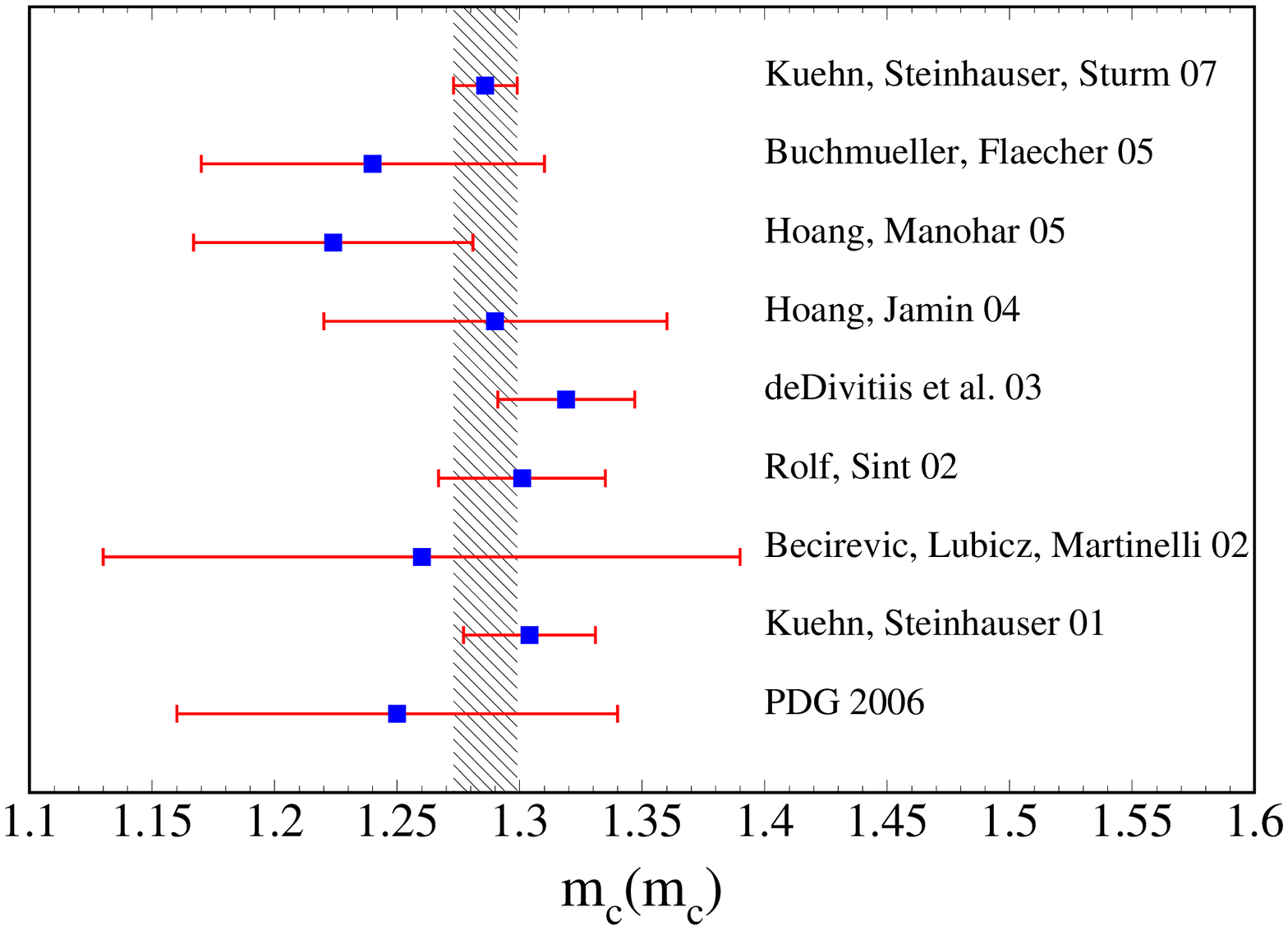}
    \end{tabular}
  \end{center}
  \vspace*{-2em}
  \caption{
    \label{fig::mc_compare}
    Comparison of recent determinations of $m_c(m_c)$ (see also
    Tab.~\ref{tab::mc_compare}).
  }
\end{figure}

A comparison with a few selected determinations (published 2001 or
later) is shown in Tab.~\ref{tab::mc_compare} and Fig.~\ref{fig::mc_compare}.
Within their respective errors all results are consistent.
Let us also mention that a first investigation of the numerical effect of the
four-loop results for the moments has been performed in
Ref.~\cite{Boughezal:2006px} where basically the analysis of
Ref.~\cite{Kuhn:2001dm} has been adopted with updated parameters.
The result of Ref.~\cite{Boughezal:2006px} reads
$m_c(m_c)=1.295(15)$~GeV. 

\begin{table}[t]
\begin{center}
\begin{tabular}{lcl|l|l}
  \hline
  \multicolumn{3}{c|}{$m_c(m_c)$ (GeV)} & Method & Reference \\
  \hline
  1.286 &$\!\!\!\!\pm\!\!\!\!$& 0.013 & 
  low-moment sum rules, NNNLO & this work \\
  \hline
  1.24 &$\!\!\!\!\pm\!\!\!\!$& 0.07 & 
  fit to $B$ decay distribution, $\alpha_s^2\beta_0$&\cite{Buchmuller:2005zv}\\
  1.224 &$\!\!\!\!\pm\!\!\!\!$& 0.017 $\pm$ 0.054 & 
  fit to $B$ decay data, $\alpha_s^2\beta_0$ &
  \cite{Hoang:2005zw} \\ 
  1.29 &$\!\!\!\!\pm\!\!\!\!$& 0.07 & 
  NNLO moments & \cite{Hoang:2004xm} \\
  1.319 &$\!\!\!\!\pm\!\!\!\!$& 0.028 & 
  lattice, quenched & \cite{deDivitiis:2003iy} \\
  1.26 &$\!\!\!\!\pm\!\!\!\!$& 0.04$\pm$0.12 & 
  lattice, quenched & \cite{Becirevic:2001yh} \\
  1.301 &$\!\!\!\!\pm\!\!\!\!$& 0.034 & 
  lattice (ALPHA), quenched & \cite{Rolf:2002gu} \\
  1.304 &$\!\!\!\!\pm\!\!\!\!$& 0.027 & 
  low-moment sum rules, NNLO & \cite{Kuhn:2001dm} \\
  \hline
  1.25 &$\!\!\!\!\pm\!\!\!\!$& 0.09 & PDG 2006 & \cite{Yao:2006px}\\
  \hline
\end{tabular}
\caption{\label{tab::mc_compare}Predictions for $m_c(m_c)$. In the second
  column some keywords concerning the used method are given
  (NNLO: next-to-next-to-leading order;   
  NNNLO: next-to-next-to-next-to-leading order).
}
\end{center}
\end{table}

\section{\label{sec::bottom}The bottom quark mass}

\renewcommand{\arraystretch}{1.1}
\begin{table}
\begin{center}
{\footnotesize
\begin{tabular}{r|rrrrrrrrrr}
\hline
  $n$ & 
  $\bar{C}_n^{(0)}$  & $\bar{C}_n^{(10)}$ & $\bar{C}_n^{(11)}$ & 
  $\bar{C}_n^{(20)}$ & $\bar{C}_n^{(21)}$ & $\bar{C}_n^{(22)}$ & 
  $\bar{C}_n^{(30)}$ & $\bar{C}_n^{(31)}$ & $\bar{C}_n^{(32)}$ & 
  $\bar{C}_n^{(33)}$
  \\
  \hline
1&$   1.0667$&$   2.5547$&$   2.1333$&$   3.1590$&$   3.4425$&$   0.0889$&$  -7.7624$&$  -0.0599$&$   1.5851$&$  -0.0543$\\
2&$   0.4571$&$   1.1096$&$   1.8286$&$   3.2319$&$   5.0798$&$   1.9048$&$$---$$&$   4.0100$&$   7.2551$&$   0.1058$\\
3&$   0.2709$&$   0.5194$&$   1.6254$&$   2.0677$&$   4.5815$&$   3.3185$&$$---$$&$   5.6496$&$  13.4967$&$   2.3967$\\
4&$   0.1847$&$   0.2031$&$   1.4776$&$   1.2204$&$   3.4726$&$   4.4945$&$$---$$&$   3.9381$&$  17.2292$&$   6.2423$\\
\hline
\end{tabular}
}
\caption{\label{tab::cn5}Coefficients of the photon polarization
  function in the $\overline{\rm MS}$ scheme as defined in
  Eqs.~(\ref{eq::pimom}) and~(\ref{eq::cn}) for   $n_f=5$. 
  }
\end{center}
\end{table}
\renewcommand{\arraystretch}{1.0}

The same approach is also applicable to the determination of
$m_b$. The coefficients $\bar{C}_n$ are listed
in Tab.~\ref{tab::cn5}. For the bounds on $C^{(30)}_n$ we assume
\begin{eqnarray}
  \label{eq::growth_bottom}
  -8.0 \le \bar{C}_2^{(30)} \le 9.5\,,\qquad
  -8.0 \le \bar{C}_3^{(30)} \le 8.3\,,\qquad
  -8.0 \le \bar{C}_4^{(30)} \le 7.4\,,
\end{eqnarray}
where the same criteria as in the charm case have been applied.
The coefficients in Tab.~\ref{tab::cn5} determine the theoretical moments
through Eq.~(\ref{eq::Mth}) where $Q_c$ has to be replaced by $Q_b=-1/3$.
The non-perturbative terms from the gluon condensate are small and will
be neglected. We included the charm mass 
terms~\cite{Corcella:2002uu} of order $(m_c/m_b)^2$ to the moments 
$\bar{C}_n$ which 
induce a shift of about $-1$~MeV in the bottom quark mass.

\begin{table}[t]
\begin{center}
{
\begin{tabular}{c|c|c|c|c}
  \hline
  &$\Upsilon(1S)$&$\Upsilon(2S)$& $\Upsilon(3S)$&$\Upsilon(4S)$
  \\\hline
  $M_{\Upsilon}$(GeV)& 9.46030(26)& 10.02326(31)& 10.3552(5)& 10.5794(12)\\
  $\Gamma_{ee}$(keV) & 1.340(18)  & 0.612(11)   & 0.443(8)  & 0.272(29) \\
  $(\alpha/\alpha(M_{\Upsilon}))^2$ 
  & 0.932069 & 0.93099 & 0.930811 & 0.930093 \\
  \hline
\end{tabular}
}
\caption{\label{tab::ResUps} Masses, electronic
 widths and effective electromagnetic
 couplings~\cite{Teubner:priv_com} 
 at $s= M^2_\Upsilon$
 for the narrow $\Upsilon$-resonances.} 
\end{center}
\end{table}

\begin{figure}[t]
  \begin{center}
    \begin{tabular}{c}
      \leavevmode
      \epsfxsize=.8\textwidth
      \epsffile{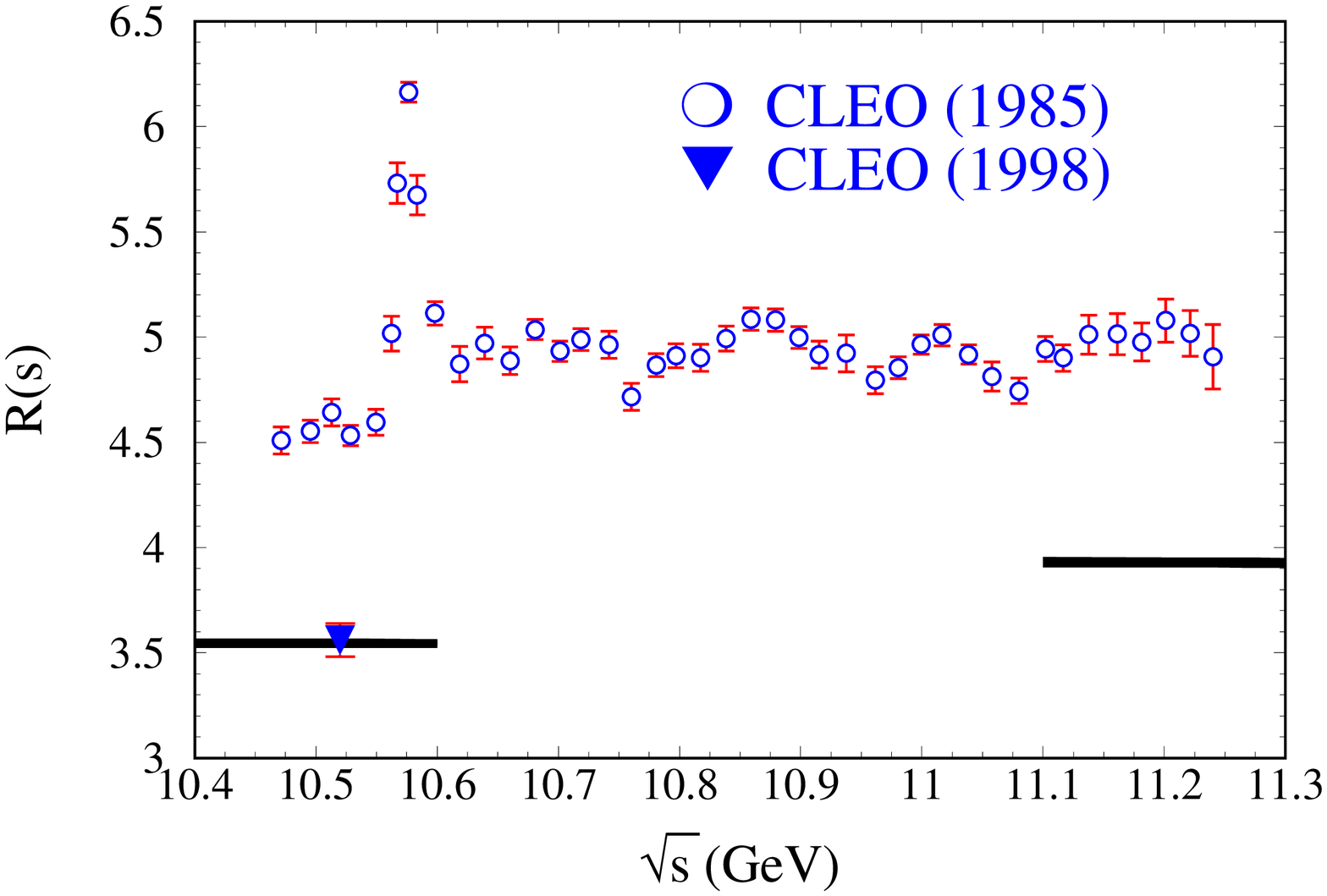}
      \\[-2em]
      \leavevmode
      \epsfxsize=.8\textwidth
      \epsffile{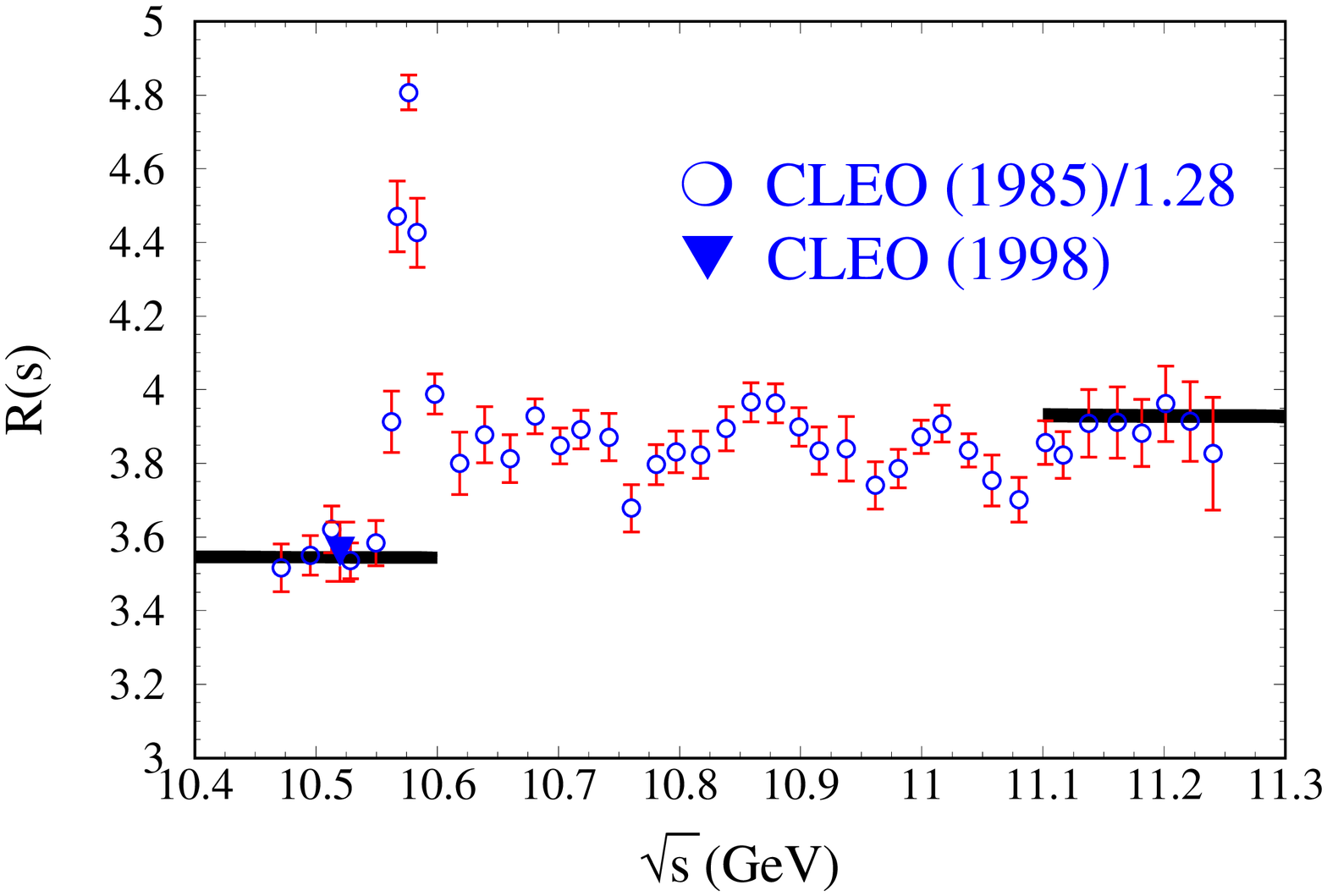}
      \\[-2em]
    \end{tabular}
  \end{center}
  \caption{\label{fig::cleodata}
    In the upper plot the data for $R(s)$ are shown as published in
    Refs.~\cite{Besson:1984bd} (circles) and~\cite{Ammar:1997sk}
    (triangle). The black curves are the predictions from pQCD outside
    the resonance region. In the lower plot the older data
    from~\cite{Besson:1984bd} are rescaled by a factor 1/1.28.
          }
\end{figure}

The experimental results for the moments are listed in Tab.~\ref{tab::Mn5}. 
The contribution from the resonances include $\Upsilon(1S)$ up to
$\Upsilon(4S)$. 
The values for the masses of the $\Upsilon$~resonances and their
electronic width have been taken from Ref.~\cite{Yao:2006px} and 
are listed in Tab.~\ref{tab::ResUps}.
The errors from the three lowest $\Upsilon$ resonances have been
combined linearly, since the PDG values for the electronic widths are
dominated by the measurement from CLEO~\cite{Rosner:2005eu}. 
The result is then combined in quadrature with the contribution from 
the $\Upsilon(4S)$ resonance.
The treatment of the bottom threshold region is similar to the one
of the charm region.
Measurements of $R$ from threshold up to 11.24~GeV have been
performed by the CLEO Collaboration more than 20 years
ago~\cite{Besson:1984bd}, with a systematic error of 6\%. 
No radiative
corrections were applied. The average value derived from the four data
points below threshold amounts to $\bar R = 4.559 \pm 0.034 ({\rm stat.})$
which is 28\% larger than the prediction from pQCD. However, a later
result of CLEO~\cite{Ammar:1997sk} at practically the same energy,
$R(10.52~\mbox{GeV})=3.56\pm0.01\pm0.07$, is
significantly more precise and in perfect agreement with theory.
Applying a rescaling factor of $1/1.28$ 
to the old CLEO data not
only enforces agreement between old and new CLEO data and pQCD 
in the region below the $\Upsilon(4S)$, it leads, in addition, 
also to excellent agreement between theory and experiment above 
threshold around 11.2~GeV where pQCD
should be applicable also to bottom production.
Further support to our approach is provided by the CLEO measurement
of the cross section for bottom quark production at
$\sqrt{s}=10.865$~GeV which is given by
$\sigma_b(\sqrt{s}=10.865~\mbox{GeV})=0.301\pm0.002\pm0.039$~nb~\cite{Huang:2006em}.   
The central value can be converted to $R_b(10.865~\mbox{GeV})=0.409$.
On the other hand, if one extracts $R_b(10.865~\mbox{GeV})$ from the
rescaled CLEO data from 1984~\cite{Besson:1984bd} one obtains
$R_b(10.865~\mbox{GeV})=0.425$ which deviates by less than 4\% from
the recent result~\cite{Huang:2006em}.
In Fig.~\ref{fig::cleodata} the original and the rescaled data
from~\cite{Besson:1984bd} is shown and compared to pQCD and 
data point from~\cite{Ammar:1997sk}.
We thus extract the threshold contribution to the moments from the interval 
10.62~GeV $\le \sqrt{s} \le$ 11.24~GeV 
by applying the rescaling factor to the
data, subtract the ``background'' from $u$, $d$, $s$ and $c$ quarks and
attribute a systematic error of 10\% to the result.  

To evaluate the contributions to the moments from
above $11.24$~GeV, i.e. ${\cal M}_n^{\rm cont}$,
we use the prediction from pQCD for $R_b(s)$.

\begin{table}[t]
\begin{center}
{
\begin{tabular}{l|lll|l}
\hline
$n$ & ${\cal M}_n^{\rm res,(1S-4S)}$
& ${\cal M}_n^{\rm thresh}$
& ${\cal M}_n^{\rm cont}$
& ${\cal M}_n^{\rm exp}$
\\
  & $\times 10^{(2n+1)}$
& $\times 10^{(2n+1)}$
& $\times 10^{(2n+1)}$
& $\times 10^{(2n+1)}$
\\
\hline
$1$&$   1.394(23)$ &$   0.296(32)$ &$   2.911(18)$ &$   4.601(43)$ \\
$2$&$   1.459(23)$ &$   0.249(27)$ &$   1.173(11)$ &$   2.881(37)$ \\
$3$&$   1.538(24)$ &$   0.209(22)$ &$   0.624(7)$  &$   2.370(34)$ \\
$4$&$   1.630(25)$ &$   0.175(19)$ &$   0.372(5)$  &$   2.178(32)$ \\
\hline
\end{tabular}
}
\caption{\label{tab::Mn5}Moments for the bottom quark system.
}
\end{center}
\end{table}

The results for the moments are listed in Tab.~\ref{tab::Mn5}, 
those for $m_b(10~\mbox{GeV})$ and $m_b(m_b)$  in Tab.~\ref{tab::mb}.
Just as in the charm case, a remarkable consistency and stability is
observed. For $n=1$ the error is dominated by the experimental input.
For $n=3$ we obtain $\pm 0.010$ from the experimental input,
$\pm 0.014$ from $\alpha_s$ and $\pm 0.006$ from the variation of $\mu$.

\begin{figure}[t]
  \begin{center}
    \begin{tabular}{c}
      \leavevmode
      \epsfxsize=0.95\textwidth
      \epsffile{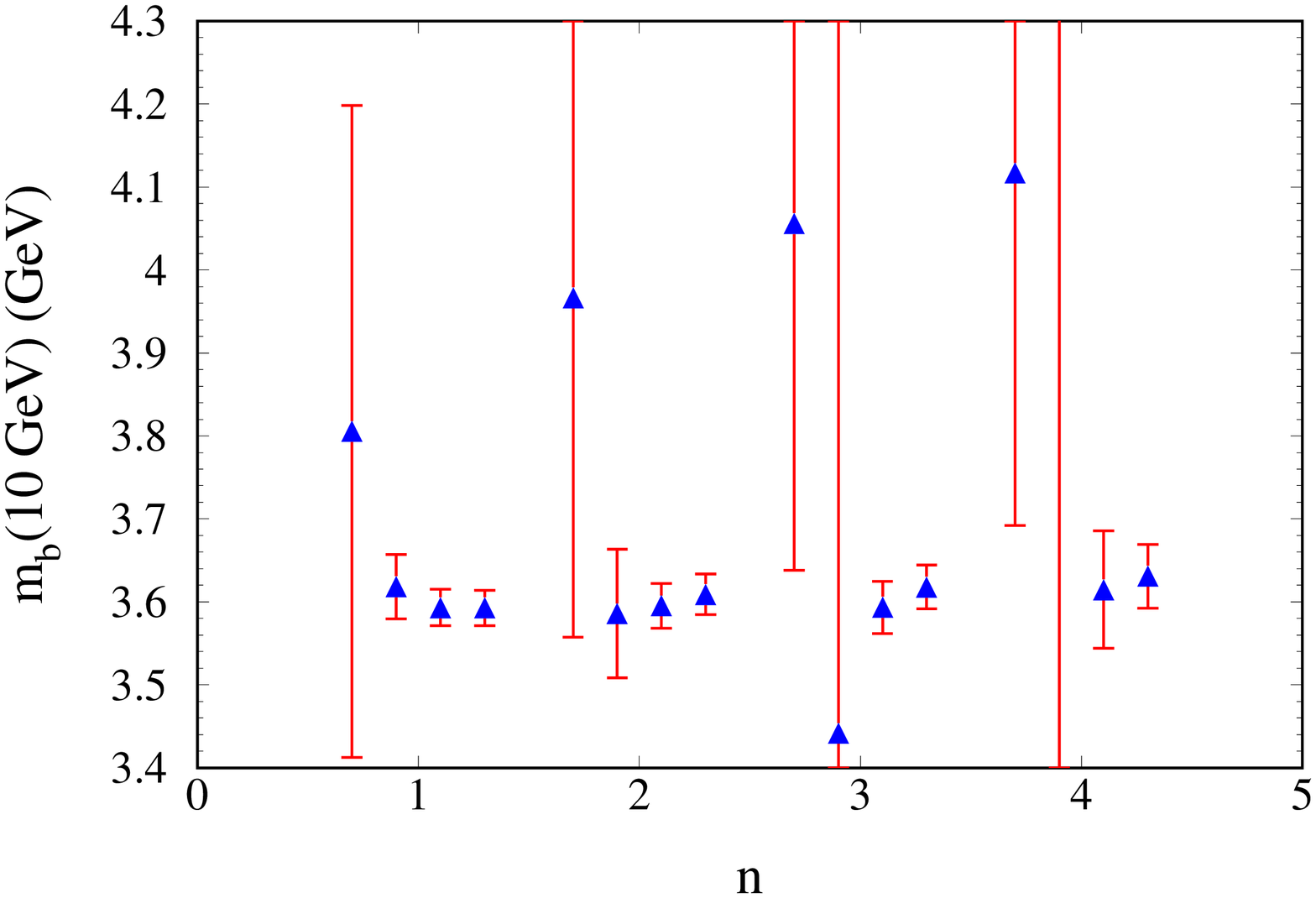}
    \end{tabular}
  \end{center}
  \vspace*{-2em}
  \caption{\label{fig::mb10}$m_b(10~\mbox{GeV})$ for $n=1,2,3$ and $4$.
  For each value of $n$ the results from left to right correspond
  the inclusion of terms of order $\alpha_s^0$, $\alpha_s^1$,
  $\alpha_s^2$ and $\alpha_s^3$ in the coefficients $\bar{C}_n$
  (cf. Eq.~(\ref{eq::cn})).
  Note, the central value for $n=4$ and the uncertainties for 
  $n=3$ and $n=4$ can not be determined
  in the case where only the two-loop corrections of order $\alpha_s$ are
  included into the coefficients $\bar{C}_n$ as the corresponding
  equation cannot be solved for $m_b(10~\mbox{GeV})$.
  }
\end{figure}

The sensitivity to the inclusion of higher orders is displayed in 
Fig.~\ref{fig::mb10}.
For the lowest moment the error is dominated by the experimental
uncertainty --- nevertheless the theory error is reduced also in this
case and the prediction stabilized. In general
a significant improvement and stabilization is observed.

The three results based on $n=1,2$ and 3 are of comparable
precision. The relative size of the contributions from the threshold
and the continuum region decreases for the moments $n=2$ and 3. On the
other hand, the theory uncertainty estimated from the variation of
$\mu$ and $\delta C_n^{(30)}$ is still acceptable. We therefore take
the result from $n=2$ (which is roughly between the $n=1$ and $n=3$
values) and add the uncertainty from ``total''
(cf. Tab.~\ref{tab::mb}) and the one 
induced by $\delta C_n^{(30)}$ linearly which leads to 
an error estimate of $\pm25$~MeV:
\begin{eqnarray}
  m_b(10~{\rm GeV} ) &=& 3.609(25)~\mbox{GeV}\,,\\
  m_b(m_b) &=& 4.164(25)~\mbox{GeV}\,.
  \label{eq::mbten}
\end{eqnarray}
This result can also be converted to a pole mass 
\cite{Chetyrkin:1999ys,Chetyrkin:1999qi,Melnikov:2000qh} of 
$M_b^{\rm(3-loop)} = 4.800~\mbox{GeV}$.
Again we do not display the additional error from the
$\overline{\rm MS}$-pole-mass conversion, which is significantly larger.

\begin{table}[t]
\begin{center}
{
\begin{tabular}{l|l|lll|l|l|l}
\hline
$n$ & $m_b(10~\mbox{GeV})$ & 
exp & $\alpha_s$ & $\mu$ &
total & $\delta\bar{C}_n^{(30)}$ &
$m_b(m_b)$ 
\\
\hline
        1&  3.593&  0.020&  0.007&  0.002&  0.021&---&  4.149 \\
        2&  3.609&  0.014&  0.012&  0.003&  0.019&  0.006&  4.164 \\
        3&  3.618&  0.010&  0.014&  0.006&  0.019&  0.008&  4.173 \\
        4&  3.631&  0.008&  0.015&  0.021&  0.027&  0.012&  4.185 \\
\hline
\end{tabular}
}
\caption{\label{tab::mb}Results for $m_b(10~\mbox{GeV})$ in GeV
  obtained from Eq.~(\ref{eq::mc1}) (adopted to the bottom quark case).
  The errors are from experiment, $\alpha_s$ and the variation of $\mu$.
  The error from the yet unknown four-loop term is kept separate.
  Last column: central values are shown for $m_b(m_b)$.
}
\end{center}
\end{table}

\begin{figure}[t]
  \begin{center}
    \begin{tabular}{c}
      \leavevmode
      \epsfxsize=0.95\textwidth
      \epsffile{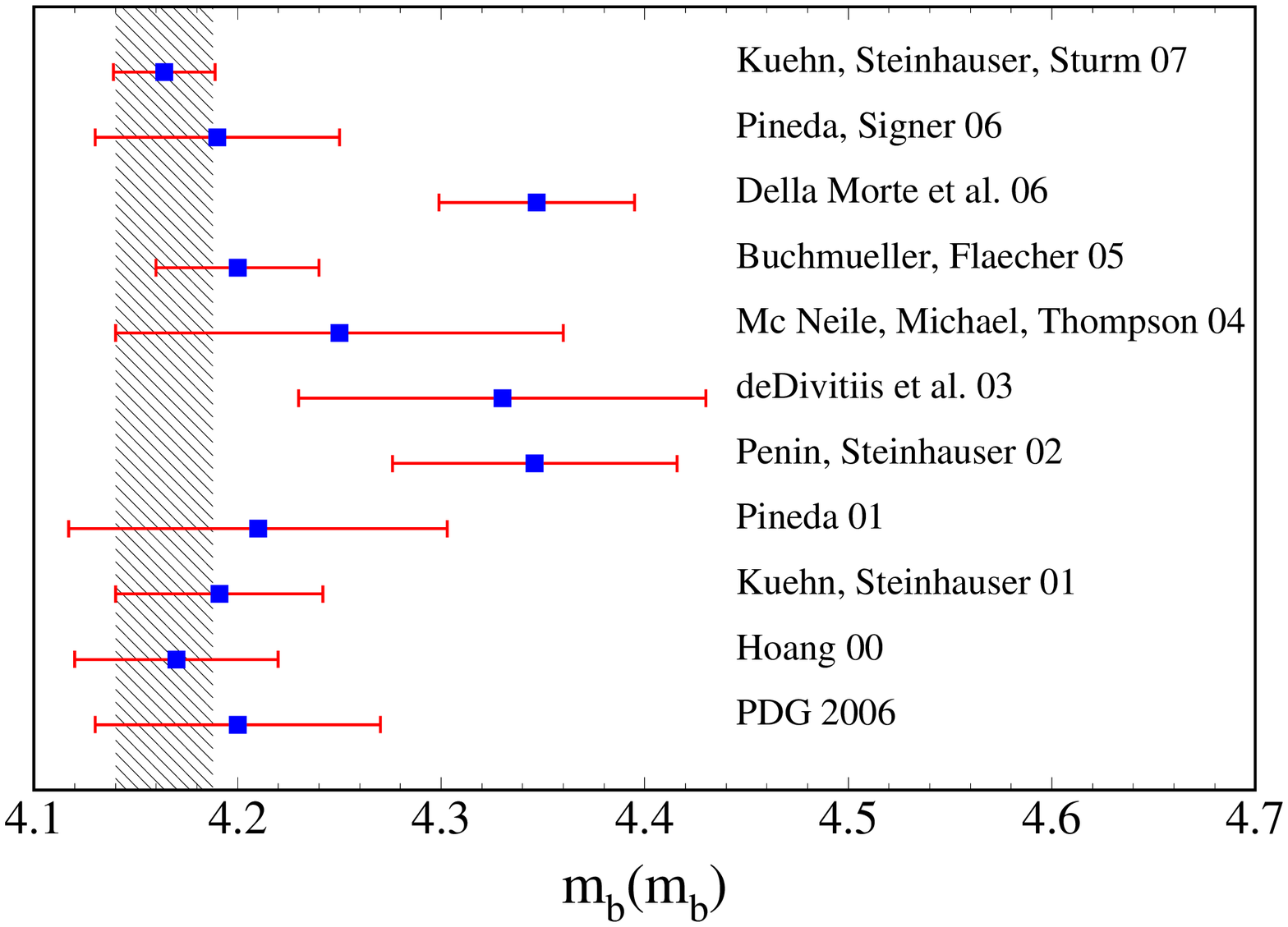}
    \end{tabular}
  \end{center}
  \vspace*{-2em}
  \caption{
    \label{fig::mb_compare}
    Comparison of recent determinations of $m_b(m_b)$ (see also
    Tab.~\ref{tab::mb_compare}).
  }
\end{figure}

A comparison with a few selected determinations
is shown in Tab.~\ref{tab::mb_compare} and Fig.~\ref{fig::mb_compare}.
In Ref.~\cite{Boughezal:2006px} $m_b(m_b)=4.205(58)$~GeV was obtained
on the bases of the four-loop result for $n=1$ and the
phenomenological analysis of Ref.~\cite{Kuhn:2001dm} with 
updated resonance parameters.

\begin{table}[t]
\begin{center}
\begin{tabular}{lcl|l|l}
  \hline
  \multicolumn{3}{c|}{$m_b(m_b)$ (GeV)} & Method & Reference \\
  \hline
  4.164 &$\!\!\!\!\pm\!\!\!\!$& 0.025 & 
  low-moment sum rules, NNNLO & this work \\
  \hline
  4.19 &$\!\!\!\!\pm\!\!\!\!$& 0.06 & 
  $\Upsilon$ sum rules, NNLL (not complete) & \cite{Pineda:2006gx} \\
  4.347 &$\!\!\!\!\pm\!\!\!\!$& 0.048 & 
  lattice (ALPHA), quenched & \cite{DellaMorte:2006cb} \\
  4.20 &$\!\!\!\!\pm\!\!\!\!$& 0.04 & 
  fit to $B$ decay distribution, $\alpha_s^2\beta_0$&\cite{Buchmuller:2005zv}\\
  4.25 &$\!\!\!\!\pm\!\!\!\!$& 0.02 $\pm$ 0.11 & 
  lattice (UKQCD) & \cite{McNeile:2004cb} \\
  4.33 &$\!\!\!\!\pm\!\!\!\!$& 0.10 & 
  lattice, quenched & \cite{deDivitiis:2003iy} \\
  4.346 &$\!\!\!\!\pm\!\!\!\!$& 0.070 & 
  $\Upsilon(1S)$, NNNLO & \cite{Penin:2002zv} \\
  4.210 &$\!\!\!\!\pm\!\!\!\!$& 0.090 $\pm$ 0.025 & 
  $\Upsilon(1S)$, NNLO & \cite{Pineda:2001zq} \\
  4.191 &$\!\!\!\!\pm\!\!\!\!$& 0.051 & 
  low-moment sum rules, NNLO & \cite{Kuhn:2001dm} \\
  4.17 &$\!\!\!\!\pm\!\!\!\!$& 0.05 & 
  $\Upsilon$ sum rules, NNLO & \cite{Hoang:2000fm} \\
  \hline
  4.20 &$\!\!\!\!\pm\!\!\!\!$& 0.07 & PDG & \cite{Yao:2006px}\\
  \hline
\end{tabular}
\caption{\label{tab::mb_compare}Predictions for $m_b(m_b)$. In the second
  column some keywords concerning the used method are given
  (NNLO: next-to-next-to-leading order;
  NNNLO: next-to-next-to-next-to-leading order; 
  NNLL: next-to-next-to-leading-logarithmic order).
}
\end{center}
\end{table}

For various applications, either related to $Z$-boson decays or in
connections to Grand Unified Theories (GUTs) the value of $m_b(\mu)$
at high scales is of interest. The result for $\mu=M_Z$ can be obtained
in a straightforward way
\begin{eqnarray}
  m_b(M_Z) &=& 2.834 \pm 0.019 \pm 0.017
  \,,
\end{eqnarray}
where the first error originates from the combined error listed in 
Eq.~(\ref{eq::mbten}), the second error from $\delta\alpha_s$.
(We do not quote a theory error at this point.)

In connection with Yukawa unification in Grand Unified Theories 
the value $m_b(\mu=m_t)$
in the $n_f=6$ theory may be of interest. Given
$M_t=171.4\pm2.1$~\cite{Group:2006xn}, 
which we identify within the present uncertainties with the pole mass,
we find 
\begin{eqnarray}
  m_t(m_t) &=& 161.8 \pm 2.0 \pm 0.2
  \,,
\end{eqnarray}
to three-loop
accuracy~\cite{Chetyrkin:1999ys,Chetyrkin:1999qi,Melnikov:2000qh}. 
(Again we do not specify a theory uncertainty. The difference between
the two- and three-loop result amounts to 426~MeV.)
Evolving $m_b$ from $\mu=10$~GeV to $m_t(m_t)$ and matching to the
$n_f=6$ theory we find
\begin{eqnarray}
  m_b(161.8) &=& 2.703 \pm 0.018 \pm 0.019
  \,,
\end{eqnarray}
where the first error reflects the combined error from
Eq.~(\ref{eq::mbten}) and the second one the uncertainty due to $\alpha_s$.
As stated above, the ratio
\begin{eqnarray}
  \frac{m_t(m_t)}{m_b(m_t)} &=& 59.8 \pm 1.3
  \,,
\end{eqnarray}
should be a useful input for Grand Unified Theories.


\section{\label{sec::con}Conclusions}

A new determination of the charm and bottom quark masses has been
presented. It is based on new experimental information on the
electronic width of the narrow quarkonium resonances and on the cross
section for charm and bottom production in the respective threshold region.
In addition the analysis profits from the improved determination of
$\alpha_s$ from a new analysis of recent experiments. Four-loop
results for the moments lead to a significant reduction of the theory
error. The result can be directly expressed in terms of running
$\overline{\rm MS}$ masses conveniently expressed at a scale of 3~GeV
and 10~GeV, respectively. Our final results
$m_c(3~\mbox{GeV})=0.986(13)$~GeV 
and
$m_b(10~\mbox{GeV})=3.609(25)$~GeV, 
correspond to 
$m_c(m_c)=1.286(13)$~GeV 
and
$m_b(m_b)=4.164(25)$~GeV.
These values are consistent with our previous determination~\cite{Kuhn:2001dm}
but considerably more precise. 
Let us stress that, since $m_c$ is relatively small,
$m_c(m_c)$ shows a less stable behaviour against higher order
corrections. Thus we propose to use $m_c(3~\mbox{GeV})$ as the better
choice for comparisons of the various methods.
The evaluation of the running bottom
mass at the scale of $m_t$ leads to 
$m_t(m_t)/m_b(m_t) = 59.8 \pm 1.3$, a result of interest for theories
with Yukawa unification.
We want to mention that for the charm quark mass our analysis
constitutes the only one where NNNLO corrections from theory are incorporated.
In the case of the bottom quark there is only one further NNNLO
analysis which is based on the mass of the $\Upsilon(1S)$ system.
However, the accuracy of this approach is strongly limited due to large
non-perturbative contributions.


\section*{Acknowledgments}

We would like to thank Thomas Teubner for providing up-to-date values
for the electromagnetic coupling at the masses of the narrow
resonances. Furthermore we thank Marco Schreck for providing the input
for $\delta R^{(3)}_{uds(c)}$.
We thank Andr\'e Hoang,
Konstantin Chetyrkin and Alexander Penin for useful discussions and 
carefully reading the manuscript.
The work of C.S. was partially supported by MIUR under contract
2006020509$\_$004 and by the European Community's Marie-Curie Research
Training Network under contract MRTN-CT-2006-035505 `Tools and Precision
Calculations for Physics Discoveries at Colliders'.
This work was supported by the DFG through SFB/TR~9.




\end{document}